\documentclass[journal,comsoc]{IEEEtran}
\usepackage{amssymb}
\usepackage{epsfig}
\usepackage{amsmath}
\usepackage{graphicx}
\usepackage{wrapfig}
\usepackage{epstopdf}
\usepackage{pifont}
\usepackage{verbatim}
\usepackage{arevmath}    
\usepackage{balance}
\usepackage{float} 
\usepackage{caption}
\usepackage{tabularx}
\usepackage{color} 
\usepackage{siunitx}
\usepackage{lettrine}
\usepackage{subcaption}
\usepackage{tikz}
\usepackage{multirow}
\usepackage{url}
\graphicspath{Figures/}
\usepackage[linesnumbered,ruled]{algorithm2e}
\SetKwComment{Comment}{/* }{ */}
\usepackage{array}

\begin{document}

\title{Content Caching Methods in Named Data Networks}

\vspace{-0.5cm}
\author{
    \IEEEauthorblockA{Pankaj Chaudhary*, Neminath Hubballi, Sameer G. Kulkarni} 
    
    \thanks{*Corresponding author: phd2001201004@iiti.ac.in \\
    Pankaj Chaudhary and Neminath Hubballi are with the Department of Computer Science and Engineering, Indian Institute of Technology Indore, India. \\
    Sameer Kulkarni is with the Department of Computer Science and Engineering, Indian Institute of Technology Gandhinagar, India.}
}

\maketitle

\begin{abstract}
Information Centric Networking (ICN) is a new network architecture (Internet) that focuses on content rather than the end-hosts. 
Named Data Networking (NDN) is a specific implementation of ICN, which relies on the use of named data and a request-response model for content distribution. These Internet architectures are known for their ability to cache content at the network level. Many caching techniques have been designed as part of various ICN/NDN projects. Caching techniques help improve the content delivery performance by storing content in the router to meet future demand. In this survey, we provide a structured review of caching algorithms designed for ICN, with a particular emphasis on NDN. We first present a taxonomy of caching techniques, followed by a detailed discussion of the various methods. Alongside their working principles, we also summarize their advantages and disadvantages. Finally, we discuss the performance metrics commonly used in the literature to evaluate caching methods and outline directions for future research in this area.
\end{abstract}


\section{Introduction}
\label{intro} 
Internet has become an integral part of our daily life. We use it for activities ranging from ecommerce, banking, social media, entertainment, etc. Internet the way it works today is designed 50 years ago with few use cases in mind and heavily influenced by the way computers worked in that period. For \textit{e.g} mobility was never a design goal. However, in the last five decades, Internet has witnessed several new applications and this coupled with developments in the transmission technologies have lead to widespread changes in the usage patterns.  Today we have very rich multimedia content being consumed by the end users over the Internet. All these changing usage patterns have been addressed through a very adhoc  solutions and retrofitted in the Internet architecture. The current Internet infrastructure relies on the Transmission Control Protocol/Internet Protocol (TCP/IP) stack, which is based on a location-centric communication model where the devices communicate using IP addresses. End hosts establish connections to content providers in order to fetch content. This leads to delivery inefficiency as more number of users access the same content. Number of attempts were made to bridge this gap. Techniques like Web caching, Content Delivery Networks (CDNs), IP Anycast, IP Multicast, \textit{etc.}\cite{Saroiu2002} emerged as solutions.  However, these adhoc solutions have their own limitations. 
Hence, keeping the today's use cases in mind, Van Jacobson \cite{Jacobson2009} advocated the need for revisiting the design principles of Internet which lead to discussions on a new clean state design. The design goals of a new Internet architecture was termed as Information Centric Networking (ICN) where network emphasizes on content delivery and efficiency in transmission completely disregarding the locations. Other design goals of the perceived ICN paradigm include resiliency, inherently securing content, and also simplification of the end users operations. 

As an attempt to design new Internet architecture in line with the stated design goals of ICN, a number of research projects emerged. Noted projects include Data-Oriented Network Architecture (DONA)~\cite{Koponen2007}, Content-Centric Networking (CCN)~\cite{Jacobson2009}, Content-Centric Inter-Networking (CONET)~\cite{Detti2011}, publish/subscribe paradigm~\cite{Fotiou2012}, Network of Information (NetInf)~\cite{Dannewitz2013}, Named Data Networking (NDN)~\cite{Zhang2014}, and many others. These proposals explored different approaches for meeting the design goals of the ICN paradigm. Few are based on a clean state approach where completely new methods for communication were presented and others envisaged a coexistence of TCP/IP networks and the new network \cite{Conti2020}.  Prominent solutions which emerged as part of these new architectural designs encompassed providing  a name based routing to simplify the end user operations, replicating the content to bring delivery efficiency and providing security to the content itself alleviating the need for end-to-end security tunnel.   
 
Among all architectures, Named Data Networking emerged as a promising proposal with ongoing research. Its popularity is attributed to its simplistic communication model. In addition, researchers have developed diverse experimental platforms like NDN testbed~\cite{NDNTestbed}, simulators (ndnSIM~\cite{Mastorakis2017}), and emulators (Mini-NDN~\cite{MiniNDN}) for testing different designs for caching solutions. Firstly, NDN adopts a hierarchical naming scheme akin to the Uniform Resource Identifier (URI) representation, consisting of a sequence of named components. Second, NDN is a consumer-driven communication where it utilizes two distinct packets for content retrieval: i) Interest packet - used by the consumer (user) to initiate the requests, and ii) Data packet - cryptographically signed content by the data producer sent to fulfill the requested content of the consumer. Third, NDN adopts a stateful forwarding plane, where the forwarders maintain state for each requested content and build intelligent forwarding strategies to retrieve the associated content. The forwarders also cache the content to serve subsequent requests. 

\textsl{Motivation and Contributions:} In order to realize name-based content retrieval and caching, several methods have been developed within the NDN framework. Caching techniques \cite{Cho2012} focus on storing content strategically to meet future consumer demands, which improves content retrieval efficiency. As there are a number of such techniques and this remains an active area of research, we aim to provide a comprehensive summary of caching methods covering their pros and cons. While we primarily focus on NDN based solutions, we also cover other prominent algorithms emerged in other architectures like CCN, publish/subscribe, etc., to provide a complete coverage of the developments in the area. Our contributions in this survey are as follows.

\noindent (i) We first present the NDN architecture and compare it with the current Internet architecture.\\
(ii) We provide a broad taxonomy of caching techniques in ICN/NDN, categorizing them based on their working principles. Furthermore, we provide insights into the merits and demerits associated with each approach. \\ 
(iii) We also discuss various relevant evaluation metrics used in the ICN/NDN literature to measure the performance of caching techniques. \\
(iv) We summarize open research challenges and suggest directions for future research in the field of caching.

\textsl{Survey Methodology:} To present a comprehensive survey on caching techniques in ICN/NDN, we selected top-quality journals, conferences, and ICN workshops from various digital databases such as ACM Digital Library, IEEE Xplore, ScienceDirect, SpringerLink, and MDPI. This includes major IEEE and ACM conferences in the field of networking, such as ACM ICN, CoNEXT, SIGCOMM, INFOCOM, GLOBECOM, IFIP, etc. We also reviewed relevant RFCs (Requests for Comments) from networking research groups and NDN technical reports. In total, we examined over 120 research articles published up to 2025 on ICN/NDN caching, covering both foundational research and current trends. To identify relevant studies, we used keywords such as caching for ICN, NDN, and CCN.

 \textsl{Paper Organization:} In Section~\ref{relatedsurvveys}, we review related survey works and compare them with our study. Section~\ref{ndnoverview} outlines the design of popular ICN architectures, with a focus on the NDN paradigm. Section~\ref{ndncaching} presents a detailed discussion of existing caching techniques in ICN/NDN. Section~\ref{futuredirections} highlights research challenges related to caching in ICN/NDN and suggests directions for future work. Finally, Section~\ref{conclusion} concludes the paper.

\section{Related Surveys}
\label{relatedsurvveys} 
Caching is one of the most important features in ICN/NDN architecture. Consequently, it has garnered major attention from the ICN/NDN research community, resulting in several publications and survey articles. However, there are several gaps in the existing survey articles. Hence, in this paper, we emphasize the significance and shortcomings of relevant prior survey articles and distinguish them from our work. 

Some earlier works~\cite{Abdullahi2015, Din2018, Ioannou2016, 
Khandaker2019, Zhang2013, Zhang2015} provide an in-depth discussion of caching strategies designed for various ICN architectures. Subsequent to these surveys, several caching techniques are designed for ICN/NDN. Our article explores recent developments in caching and also provides various evaluation metrics for assessing caching performance. 

The survey in \cite{Abdullahi2015} discusses on-path and off-path caching techniques for different ICN architectures. However, there is a lack of discussion on individual caching methods. In \cite{Ioannou2016}, the authors classified on-path caching techniques into probabilistic, graph-based, label-based, and popularity-based categories. However, their study is limited to a detailed discussion of on-path caching and the challenges associated with on-path techniques. The authors of \cite{Din2018} presented a survey on content caching by classifying the existing ICN caching techniques into location-based, multi-level, single-node-based, and popularity-based categories. Khandaker \emph{et al.}~\cite{Khandaker2019} discuss notable caching techniques under each category: popularity-based, location-based, collaboration-based, and path-based. Although this survey provides a comprehensive analysis of caching in ICN, it compares the pros and cons of caching techniques using only a limited number of performance metrics (cache hit ratio and delay) for evaluation.

The authors of \cite{Alubady2023} presented a survey on modern caching techniques in NDN, classifying them into content placement and replacement methods. Although this survey provides a comprehensive analysis of caching techniques and challenges in NDN, there is a lack of discussion on cooperative and off-path caching strategies. In contrast, our survey presents a broad taxonomy of caching and covers both standard and advanced caching techniques under each category. Zhang \emph{et al.}~\cite{Zhang2023} presented a survey on content placement and replacement techniques for the ICN-based Internet of Things (IoT) domain, classifying them into conventional and machine learning-based techniques. Although the paper provides a comprehensive analysis, it lacks key evaluation metrics for ICN-IoT. The authors of \cite{Naeem2024} presented a survey on centrality-based on-path caching techniques for the NDN-based IoT domain. This survey is specific to centrality-based caching in IoT networks. However, our survey is not limited to a particular domain; in fact, we cover caching techniques for a range of application domains, including IoT, mobile networks, vehicular networks, and satellite networks.

Many articles \cite{Ahlgren2012, Fang2018, Khelifi2020, Xylomenos2013} have presented comprehensive studies on the features and functionalities of various ICN-related projects. The studies in \cite{Ahlgren2012, Xylomenos2013} offer an in-depth discussion of the features and functionalities of various ICN architectures. In \cite{Saxena2016}, the authors presented a comprehensive study on NDN architecture, services, and applications. The works in \cite{Amadeo2014, Fang2018, Khelifi2020} discuss various ICN/NDN services in the context of wireless networks. In \cite{Jmal2017b}, the authors studied different features of CCN architecture in the context of Software-Defined Networking (SDN). Zhang \emph{et~al.} \cite{Zhang2018b} provided an in-depth discussion of integrating the SDN paradigm with ICN. The surveys presented in \cite{Liang2018, Conti2020} provide guidance on working with the integration of IP architecture and ICN/NDN, offering valuable insights into the coexistence of these networking paradigms. 
However, detailed discussions on the ICN/NDN caching techniques are lacking. The primary goal of our survey is to comprehensively review the recent literature in the area of ICN and provide our insights on recent advancements in caching.
Table~\ref{tab:surveycomparison} provides a summary of existing surveys on ICN/NDN content caching techniques comparing these works to our survey.

\begin{table*}[!t]
    \centering
    \footnotesize
     \caption{Summary of Related Surveys on Caching in ICN/NDN.} 
     \label{tab:surveycomparison} 
     \resizebox{\linewidth}{!}{%
    \begin{tabular}  {|p{2.6cm}|>{\centering\arraybackslash}p{1cm}|>{\arraybackslash}p{9cm}|}
        \hline
         \textbf{Work} & \textbf{Year} & \multicolumn{1}{l|}{\textbf{Description}} \\
        \hline
        
        Zhang \emph{et~al.} \cite{Zhang2013} & 2013 & Content caching in terms of placement and replacement strategies are studied. \\
       \hline
        Zhang \emph{et~al.} \cite{Zhang2015} & 2015 & Caching strategies are categorized into six types: Homogeneous, Heterogeneous, Cooperative, Non-Cooperative, On-Path, and Off-Path Caching. \\
       \hline
        Abdullahi \emph{et~al.} \cite{Abdullahi2015} & 2015 & Discussed on-path and off-path caching techniques for various ICN architectures. \\
       \hline
        Bernardini \emph{et~al.} \cite{Bernardini2016} & 2016 & Popular caching strategies for CCN are discussed and compared. \\
       \hline
        Ioannou \emph{et~al.} \cite{Ioannou2016} & 2016 & 
       On-path caching strategies are discussed and categorized into four types: Probabilistic, Graph-Based, Label-Based, and Popularity-Based caching. \\
        \hline
        Din \emph{et~al.} \cite{Din2018} & 2017 & Caching strategies are discussed and categorized into 4 groups, viz., Location-Based, Multi-Level, Single Node-Based, and Popularity-Based caching. \\
        \hline
        Khandaker \emph{et~al.} \cite{Khandaker2019} & 2019 & Caching techniques in ICN are discussed and categorized into: Popularity-Based, Location-Based, Collaboration-Based, and Path-Based caching. \\
        \hline
        Alubady \emph{et~al.} \cite{Alubady2023} & 2023 & Caching strategies for NDN are discussed by categorizing them into placement and replacement techniques. \\
        \hline
        Zhang \emph{et~al.} \cite{Zhang2023} & 2023 & Surveyed conventional and machine learning caching and replacement techniques for ICN-IoT. \\ \hline
        Naeem \emph{et~al.} \cite{Naeem2024} & 2024 & Discusses different centrality-based caching techniques in ICN/NDN for IoT networks. \\
        \hline
        This work  & 2026 &  Presents a comprehensive survey of recent advancements in ICN/NDN caching by organizing existing techniques into different categories for readers to understand more easily. \\
        \hline
\end{tabular}}
\end{table*}

\section{Overview of NDN Architecture}
\label{ndnoverview}

In this section, 
we provide a brief overview of NDN architecture 
covering
NDN entities, packet structure, and the essential data structures involved in packet processing. Subsequently, we will explore the fundamental features of NDN and compare their functionalities with the existing TCP/IP architecture of the Internet.

\subsection{NDN Entities}

The NDN network comprises of three essential entities for information exchange:
\textit{i) Consumer:} This entity initiates the request for content.
\textit{ii) Producer:} An entity that generates and provides the content. 
\textit{iii) Router:} An entity tasked with forwarding packets towards potential producers to retrieve content for the requesting consumer. Additionally, routers play a crucial role in caching the requests and associated content for efficient data retrieval.

\subsection{NDN Packet}
\begin{figure}[!hthp]
\centering
\begin{subfigure}{0.2\textwidth}
    \vspace{-2mm}
   \scalebox{0.8}{\includegraphics{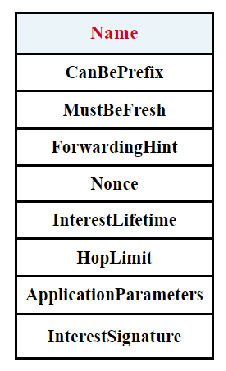}}
     \caption{Interest Packet}
      \label{fig:interest}
\end{subfigure}
\begin{subfigure}{0.2\textwidth}
    \vspace{-2mm}
   \scalebox{0.8}{\includegraphics{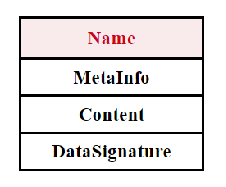}}
     \caption{Data Packet}
     \label{fig:data}
\end{subfigure}
\vspace{-2mm}
 \caption{NDN Packet Structure \cite{ndn-packet-spec}}
  \label{fig:ndn_packet_format}
\vspace{-6mm}
\end{figure}

NDN communication relies on two types of packets, namely Interest (Request) and  Data (Content) packets~\cite{ ndn-packet-spec, Zhang2014}. These packets enable information exchange between the consumers and producers in the network. Fig.~\ref{fig:ndn_packet_format} illustrates the structure of NDN packets.
\textit{i) Interest:} Consumers use Interest packets to indicate their desire for a specific piece of content identified by a unique name. These packets travel through the network router towards the potential source of the requested content.
\textit{ii) Data:} Data packets serve as responses to the Interest packets, which contain the requested content from the producer. The routers, while forwarding the Data packets towards the consumer may cache the content locally according to the caching policy of the network.

\subsection{NDN Packet Forwarding Data Structures}
\begin{figure*}[!h]
\centering
  \includegraphics{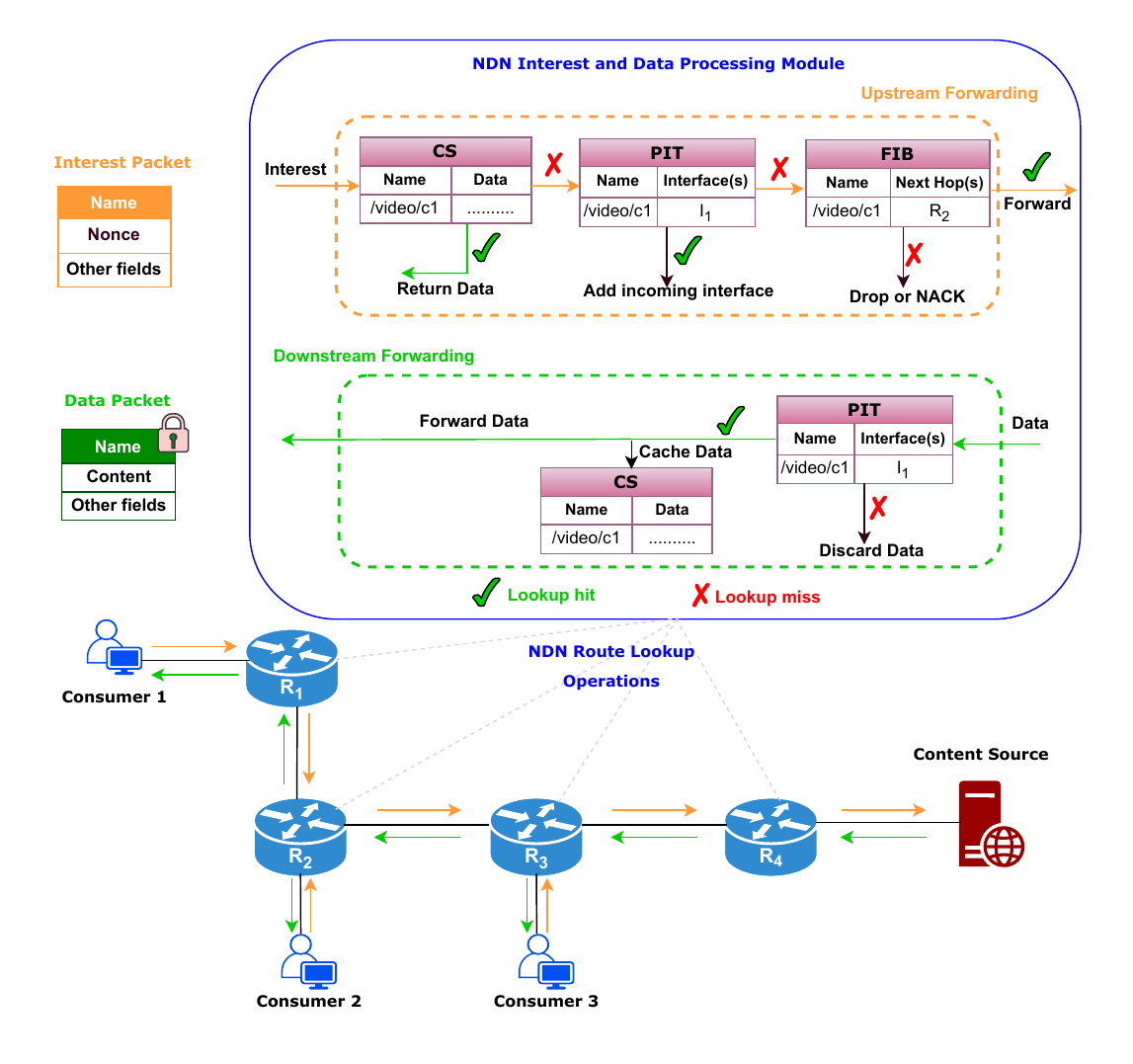}
  \caption{NDN Packet Processing through NDN Router}
  \label{fig:NDNArchitecture}
\end{figure*}

NDN routers have three data structures for packet forwarding, namely, \textit{i)} Content Store (CS), \textit{ii)} Pending Interest Table (PIT), and \textit{iii)} Forwarding Information Base (FIB). These data structures are built using application names rather than IP addresses \cite{Afanasyev2018}. 
Fig.~\ref{fig:NDNArchitecture} depicts how the NDN router uses these three data structures to process NDN packets.

\textbf{Content Store:} The NDN routers cache the received content at CS for future reference. When a consumer requests for data, an Interest packet is sent upstream towards the data source, where the intermediate NDN routers check whether the data exists in their CS. If the requested data exists, then they send it to the consumer in the reverse direction. But, when the requested data is not found in their CS, the routers proceed with a PIT lookup.

\textbf{Pending Interest Table:} The PIT entries include names of all the interface(s) which have pending Interest(s).  These interest requests have been forwarded but have not yet been serviced. 
When the lookup in CS fails, the NDN router checks its PIT. If a PIT entry exists for the requested Interest, then it adds the incoming interface number/name to the PIT provided the interface is not already marked and then it drops the Interest. This indicates that the request has already been forwarded to fetch the content. 
Otherwise, the request is passed to the FIB, and the outgoing interface in PIT is marked. In the downstream path, \textit{i.e.}, after receiving the Data packet from the producer or the neighboring upstream router, the NDN router checks its PIT entry to find the matching interfaces, and forward the Data packet to all the listed interfaces. The router then removes the entry from the PIT and may cache the Data packet in its local CS. However, when there is no match in the PIT for the corresponding name prefix, the Data packet is dropped. This means that the router has either already served the content in response to the requested Interest or the entry has become stale.

\textbf{Forwarding Information Base:} Every NDN router maintains a FIB that contains name prefixes and associated outgoing interface(s) in order to forward Interest(s) to the content producer(s). The name-based routing protocol is used to populate the FIB entry. 
FIB in NDN lists all potential interfaces for each name prefix and ranks interfaces according to the forwarding policy to determine the best pathways for packet delivery.

\subsection{Fundamental Features of NDN}
The main features of NDN architecture are the following. 
\paragraph{\textbf{Naming}} NDN operates on a content-centric model, prioritizing the retrieval and distribution of content over the establishment of connections between devices. Unlike IP networks, NDN focuses on naming the content rather than addressing the location of the content source for routing packets. Each data unit in the network is identified by a unique name, and the consumers request data by specifying this name. 
NDN routers forward packets and cache content based on the content name carried in the packets. 
NDN adopts a hierarchical naming convention of unbounded lengths comprising multiple components separated by a delimiter (`/') \cite{NDN2021}. For example, \texttt{/abc/xyz/Name.mp4}. These names are like Uniform Resource Locators (URLs) in a human-readable format. 

\paragraph{\textbf{Routing and Forwarding}} Routing in NDN involves determining the most efficient paths for routing Interest packets to potential content providers and Data packets to nodes making requests. 
Routing protocols populate the FIB of each router with multiple next-hop interfaces corresponding to specific name prefixes. Based on this information the routers route the request towards the potential content source. The router may direct the request to either one of the interface or multiple interfaces. If a router has the requested content in its CS for the specified name prefix, it responds with a Data packet, following the reverse path of the corresponding Interest packet. 

\paragraph{\textbf{Caching}} Content caching plays a pivotal role in NDN. It empowers the routers to store and serve the content to consumers with improved efficiency facilitating faster data retrieval, efficient bandwidth utilization, and high availability of data. 
Caching in NDN can be performed in two distinct forms: On-path and Off-path. In On-path caching, the content is cached at the routers situated on the transmission path between the producer and consumer. On the other hand, Off-path caching provides the flexibility to cache content at any location within the network. 
The caching decision involves choosing what content to cache, where to cache it, and also determining which content to evict when the cache is full. The decision-making process for content caching by the router can occur either cooperatively \textit{i.e,} routers consult with neighbors to make caching decisions or non-cooperatively, where each routers independently decide whether to cache the content or not. Cooperative decisions maximize cache space utilization by reducing content redundancy, while the non-cooperative approach minimizes communication burdens on routers.


\paragraph{\textbf{Security}} In NDN, security is inherent in the architecture and provides data integrity, confidentiality, and authenticity \cite{Zhang2018c} to content. 
This is achieved through a combination of digital signatures, encryption and trust models \cite{Afanasyev2021}. 
NDN integrates security features at the network layer, ensuring verifiable and trustworthy communication. Producers sign each Data packet based on the namespace to guarantee data integrity and origin authentication. 
This approach enhances content delivery security and mitigates diverse security threats.

\subsection{IP vs. NDN: A Comparison}
\begin{figure}[!bthp]
\scalebox{0.5}{\includegraphics{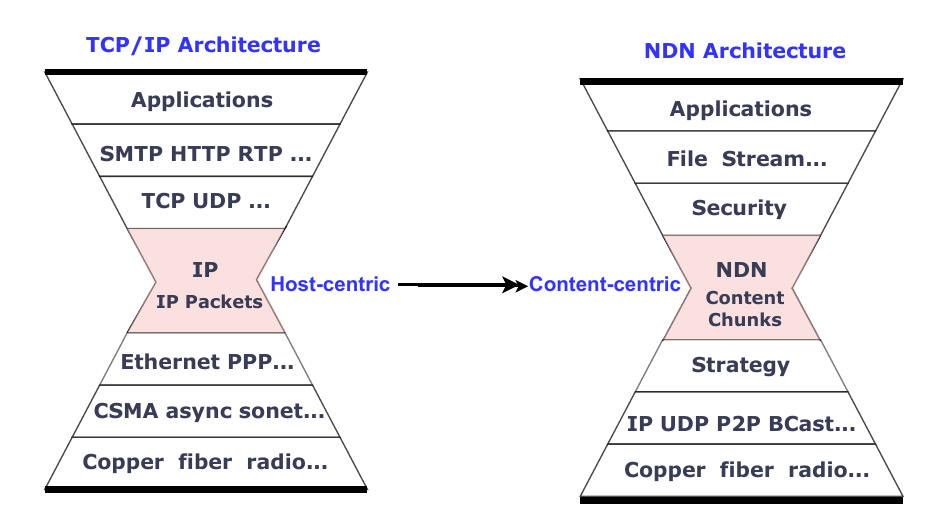}}
  \caption{Hourglass Models of TCP/IP and NDN Protocol Stacks} 
  \label{fig:ipvsndn}
  \vspace{-4mm}
\end{figure}

Here, we present a comparison of the IP and NDN architectures.  
Fig.~\ref{fig:ipvsndn} shows the hourglass model of Internet architecture, depicting traditional IP and NDN networks. 
We can see that NDN adopts the hourglass shape of the TCP/IP architecture by modifying the narrow waist of the IP network layer \cite{Zhang2014}. In IP, each communication endpoint is identified by an IP address to enable communication and data exchange. In contrast, the narrow waist of NDN comprises of content names, where data is requested and retrieved by its name rather than its location.
Table~\ref{tab:comparisonIP_NDN} offers a comprehensive comparison of the functional dissimilarities and commonalities between IP and NDN paradigms.
\begin{table*}[bthp]
\centering
\caption{Comparison Between IP and NDN}
\label{tab:comparisonIP_NDN}
\footnotesize
\begin{center}
\resizebox{\linewidth}{!}{%
 \begin{tabular}{|p{0.20\linewidth}|p{0.55\linewidth}|p{0.55\linewidth}|} 
 \hline
 \centering{\textbf{Features}} & \centering{\textbf{IP}} & \centering{\textbf{NDN}}  \tabularnewline 
\hline

Communication Model &  Host-centric (location-dependent)  & Content-centric (location-independent) \\
 \hline
 
Namespace & 
 
\textbullet \hspace{0.5em} \parbox[t]{\dimexpr\linewidth-2.0em}{Routes packets using IP addresses}

\textbullet \hspace{0.5em} Namespace is bounded by the number of IP addresses. 

\textbullet \hspace{0.5em} \parbox[t]{\dimexpr\linewidth-2.0em}{Application names convert to IP addresses for network use} 
& 
\textbullet \hspace{0.5em} Uses hierarchical structured names 

\textbullet \hspace{0.5em} Theoretically unbounded/infinite namespace.

\textbullet \hspace{0.5em} \parbox[t]{\dimexpr\linewidth-2.0em}{No conversion needed; network uses application name directly} \\ \hline

 Routing/Control Plane &
  
\textbullet \hspace{0.5em} Stateful and adaptive

\textbullet \hspace{0.5em} More load on the routing plane
 &
 \textbullet \hspace{0.5em} Stateful and adaptive
 
\textbullet \hspace{0.5em} Reduced burden on the routing plane  \\ 
 \hline
 
Forwarding/Data Plane &
 
\textbullet \hspace{0.5em} Stateless and not adaptable

\textbullet \hspace{0.5em} Inefficient in handling network faults
&
\textbullet \hspace{0.5em} Stateful and adaptive

\textbullet \hspace{0.5em} Capable of dealing with network faults \\  \hline
 
Packet Reusability & IP packet reuse is challenging due to location dependency &
Data packets are reusable as they are location-independent \\ \hline

 Multipath & Support for multipath routing is limited & Multipath routing is fully supported  \\ \hline
 
 Caching  & Content caching in network nodes is not supported & Supports content caching at the network layer (in-network caching) \\
 \hline
 Security & Secure communication channel  &  Secure Data packet  \\
 \hline
 Mobility  & Lack of mobility support & Efficient mobility support \\
 \hline
 Efficiency  & Less efficient for content retrieval and distribution & More efficient for content retrieval and distribution \\
 \hline
 Application & Primarily used for network communication and Internet browsing &  Well-suited for content distribution \\
 \hline
 \end{tabular}}
\end{center}
\end{table*}

\section{Caching in NDN}
\label{ndncaching}

In NDN, caching stands out as an important feature that shapes the architecture's effectiveness. NDN's strength lies in improving user experience by smartly placing content at strategic locations. The caching techniques within NDN architecture deal with two crucial components: i) the Content Placement Decision and ii) the Content Replacement Decision. The former involves routers in the network making informed choices about whether or not to cache a specific piece of content. This decision is pivotal in optimizing the overall efficiency and responsiveness of the network.
On the other hand, the Content Replacement Decision is the subsequent step, where the routers determine which content to evict from their cache to make room for new content when the cache is full. This process ensures that the limited cache size on the router is constantly updated with the most relevant and in-demand content. The dynamic interaction between content placement and replacement not only optimizes resource usage, but also significantly contributes to boosting the overall efficiency and responsiveness of NDN frameworks.

We explore various caching techniques and classify them based on how they position content in the network routers. These methods can be categorized into centralized, decentralized, on-path, off-path, cooperative, non-cooperative, content popularity-aware, probabilistic, topology-aware, and content granularity-level techniques. It's worth to note that caching techniques designed for ICN/NDN often incorporate a combination of these techniques, making it challenging to assign them to specific categories. For example, popularity-based methods may employ both centralized and decentralized approaches, as well as on-path and off-path techniques.

In this section, we begin by providing a brief overview 
on the caching techniques, and 
explore the operational mechanisms of different caching schemes designed for ICN/NDN.

\subsection{Caching Techniques Taxonomy}
\begin{figure}[!bthp]
\centering
  \includegraphics[width= 0.9\columnwidth]{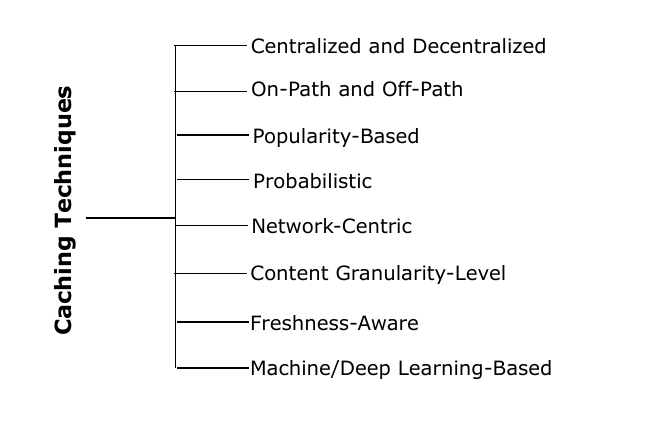}
  \caption{Taxonomy of ICN/NDN Caching}
  \label{fig:cachingtaxonomy}
  \vspace{-2mm}
\end{figure}

Based on the nature of content retrieval and caching decisions, we classify existing ICN caching techniques into various categories. The comprehensive categorization of ICN caching strategies is illustrated in Fig.~\ref{fig:cachingtaxonomy}. We first give an overview for each category and then elaborate different works taking reference to the category they belong to. Table~\ref{tab:caching_definitions} provides a summary of each category along with a brief description.


\noindent \textit{1) \textbf{Centralized and Decentralized Caching:}} 
  This approach addresses how routers in the network engage in the decision-making process to cache or not to cache the content. It explores how routers actively manage caches, searching for content, deciding on content caching based on various factors, and making room for new content when cache becomes full. The categorization encompasses various approaches, such as centralized, decentralized, cooperative, and non-cooperative techniques that we discussed below.
   
  \noindent \textbf{Centralized Caching:} Centralized caching refers to a scenario where cache storage within a network is managed in a centralized manner. In this model, a single entity, typically a centralized SDN controller \cite{Nguyen2013a, Zhang2020a}, assumes the responsibility of making caching decisions for the entire network. These caching techniques facilitate strategic content placement, enabling comprehensive network-wide optimization and efficient handling of off-path content requests. Although it enhances resource utilization efficiency, challenges such as scalability, communication overhead, and potential single point of failure are a few issues of these methods. 
   
\noindent \textbf{Decentralized Caching:} In a decentralized caching paradigm, the decision-making process is distributed across multiple routers in the network, allowing each router the autonomy to actively participate in decisions regarding content storage and retrieval \cite{Cho2012, Ren2014}. This approach empowers routers to make content caching decisions independently. Routers have the flexibility to collaborate with peers, engaging in collective decision-making processes, or routers can make decisions based on their local information. Decentralized decision-making in routers enhances adaptability to dynamic conditions, improving network resilience and scalability. However, challenges may arise in maintaining coordination among autonomously operating routers to optimize cache utilization.

 \noindent \textbf{Community-Centric Caching:} The network topology is divided into multiple communities \cite{Xu2024} to better utilize the storage capacity of routers. This division can be based on relationships among nodes or nodes with common interests, ensuring that nodes within the same community share similar characteristics \cite{Huang2019}. In this approach, nodes within a community collaborate through a leader or designated node to optimize content searching and caching decisions. The leader, elected within the community, is chosen based on its significance to the community. The community-centric approach enhances cache utilization, reduces load imbalance in networks, and minimizes content redundancy through cooperation. However, challenges arise in forming communities and selecting representative nodes (in each community) to facilitate this cooperation.

\noindent  \textbf{Cooperative Caching:} In a cooperative approach, routers within the network work together to optimize the limited cache storage \cite{Chaudhary2025, Huang2019, Li2012b}. This collaborative effort improves content retrieval efficiency as routers share partial or complete cache state information with others to help them make decisions on caching. However, this comes with the additional cost of communication. By strategically deciding on content caching, the cooperative approach contributes to the improvement of both the cache hit ratio and network-wide cache diversity (see Section~\ref{metrics} for details). Moreover, the enhancement in content availability resulting from the collaborative approach enables routers to fulfill both on-path and off-path requests. The establishment of collaboration among routers can take place through either a centralized or decentralized decision-making approach. Depending on how routers cooperate, the collaborative methods can be classified into implicit and explicit techniques \cite{Zhang2013}. 
      
   \textit{i) Implicit:} In implicit coordination, routers refrain from sharing their cache state information through extra communication messages \cite{Zhang2013, Zhang2015}. Instead, cooperation can be achieved differently, such as by appending information to Interest or Data packets during regular network interactions. This minimizes communication costs as routers share limited information, improving content availability compared to non-cooperative techniques. However, this method fails to take full advantage of the content availability in the cache of neighboring routers.
      
   \textit{ii) Explicit:} In the explicit mode of cooperation, routers explicitly share either complete cache state information or the content availability, or popularity details with other routers \cite{Fan2020, Zhang2013}. 
   Explicit cooperation proves beneficial in scenarios where a centralized controller gathers information about routers in the network or when a cluster head establishes communication within or among multiple clusters. Additionally, it is advantageous for establishing communication within the 1-hop, 2-hop, or N-hop neighborhood, as well as for cooperation among routers located on the transmission path. In explicit cooperation, routers have the flexibility to coordinate either at a local or global level with other routers in the network. Local collaboration involves coordination on the transmission path, within the neighborhood, or within specific domains. In contrast, global cooperation involves establishing coordination among routers to acquire knowledge of the entire network topology. Explicit collaboration enhances cache diversity and the availability of content for both on-path and off-path requests. However, it comes at the cost of increased communication overhead.
      
\noindent \textbf{Non-Cooperative Caching:} Here individual routers autonomously determine which content to store in their caches, what to evict, and how to route Interest and Data packets to fulfill consumer requests without coordination with other routers in the network \cite{Din2018, Zhang2017}. The simplicity of its implementation and the lower communication overhead makes it an attractive choice,  as routers need not share their cache state information with others. Despite these advantages, the non-cooperative approach encounters difficulties in attaining optimal resource utilization due to the limited cache size of the router. Routers, acting independently, may overlook the caching state of neighboring routers, resulting in content redundancy as multiple routers might cache the same content already stored by others. Moreover, the effectiveness of the non-cooperative approach declines as it overlooks broader request patterns.
  
\noindent  \textit{2) \textbf{On-Path and Off-Path Caching:}}     Caching strategies are characterized as on-path or off-path \cite{Draxler2012} based on the location of routers in the network relative to the content delivery path between the consumer and provider. Intermediate routers positioned on the transmission path between the consumer and provider are termed as on-path routers, while any other routers not directly situated on the transmission path are considered off-path routers. This classification focuses on how routers strategically position content within their caches, influencing the efficiency of content retrieval and distribution across the network.

\noindent    \textbf{On-Path Caching:} The on-path approach is inherent in the ICN/NDN architecture, where Interests are forwarded via the best path to the content source, and Data packets take the reverse path created by the Interest \cite{Zhang2014}. Routers consult their FIB to guide the Interest packet to the next hop along the path towards the potential content provider, and only routers situated along this path query for the content \cite{Psaras2012}. Once the content is found, the Data packet follows the same path from the provider to the consumer, and only the routers situated along this route are eligible for content caching \cite{Amadeo2022}. In the downstream direction, content is either cached by all routers on the path or strategically selected routers. The implementation of on-path caching is simple and can be used with both cooperative and non-cooperative setups. It reduces content access time, network traffic, and content query or caching overhead, as content can be served directly from routers on the transmission path. However, on-path techniques may face challenges with content availability as they do not leverage the caches of routers located outside the transmission path, potentially reducing the chances of finding content in cached nodes also known as cache hit ratio.
    
\noindent \textbf{Off-Path Caching:} The off-path caching techniques offer flexibility in terms of both content querying when forwarding the Interest packet towards the potential provider and caching the content in the response \cite{Draxler2012}. Off-path caching enables content to be cached beyond the transmission path, ensuring its availability for off-path requests. Additionally, routers on the forwarding path can retrieve content from the routers located outside this path, whether they are situated at a distance of 1-hop, 2-hop, or N-hop. Off-path caching can be accomplished through various methods. One approach incorporates a centralized SDN controller \cite{Nguyen2013a} with a global topology view that directs where content requests should be forwarded and cached. Alternatively, network topology partitioning into multiple domains \cite{Yan2017} and electing leader nodes to facilitate off-path caching. Another approach includes the utilization of predefined rules (e.g., content retrieval and placement using the corresponding hash value) \cite{Saino2013} or modifications to packet structures \cite{Chaudhary2023}, which can be employed to determine where Interest packets are forwarded and where Data packets are cached. Off-path caching enhances content availability across the network and allows routers to retrieve content from any neighborhood routers. However, the implementation of off-path caching is complex and introduces communication overhead to leverage its benefits.

\noindent  \textit{3) \textbf{Popularity-Based Caching:}}    Popularity of content is determined by the count of the number of times consumers demand a particular piece of content; higher the count, the more popular the content is considered \cite{Bernardini2013, Ong2014}. Unlike popularity-agnostic approaches, where routers make caching decisions without prioritizing any content, popularity-based schemes strategically cache content based on consumer demand. The content popularity metric assists routers in deciding which content to place in their caches and which to remove. By keeping frequently requested content closer to consumers, these schemes enhance performance and reduce the load on the original content source. Within the network, routers can make caching decisions based on either local popularity estimation or global estimation based on requests from all consumers. Whether content is deemed popular for caching purposes can be determined by utilizing static or dynamic threshold values. 
   
\noindent \textbf{Local Popularity:} Local popularity estimation involves analyzing consumer content access patterns within specific regions. Each router keeps a popularity table \cite{Amadeo2020, Ong2014}, monitoring requested content names and their access frequencies. When a content request arrives, the router updates this table to make informed decisions on whether to cache the content in its CS. This localized approach enables routers to customize caching decisions to meet the specific needs of regional demands, thereby enhancing content availability and user experience while minimizing coordination overhead. However, the challenge arises from the inclination towards local content preferences, potentially neglecting broader trends or globally popular content request patterns from diverse consumer regions, which could affect overall system performance.
   
\noindent \textbf{Global Popularity:} This method assesses the overall popularity of content across the entire network, improving network-wide content availability through strategic caching of globally popular content \cite{Chaudhary2025, Gui2020}. Routers store widely requested content in their cache by considering the combined demand from diverse consumer regions instead of specific ones. Global popularity estimation within the network is executed by centralized entities \cite{Zhang2020a}, which gather local popularity information from other routers. Alternatively, routers collaborate with each other or with specified routers to exchange their local popularity information to obtain a global count \cite{Gui2020, Hubballi2024b}. This process is executed either periodically at specified time intervals or after a certain number of requests or it can be performed on-demand as needed. While improving network-wide content availability, estimating global popularity comes with challenges, such as node collaboration to track global request counts and manage the popularity table. This may lead to additional communication and computational overhead. In large-scale networks with diverse user behaviors, the efficiency of the caching system may confront subtle challenges.
   
\noindent \textbf{Static Popularity:} In this approach, content popularity is determined by analyzing historical request patterns or adhering to predefined thresholds \cite{Alduayji2023, Bernardini2013, Meng2023}. The assumption here is that the content's popularity remains consistent over a specific time period or adheres to a predictable trend. For example, caching a highly demanded video during peak periods at important network locations. In the static method, content with request frequency above the set threshold is marked as popular, while content below it is considered unpopular and may not be cached. This technique is simple and requires less effort as routers don't have to frequently modify the threshold values, making it suitable for stable content popularity trends. However, network performance diminishes as content popularity dynamically fluctuates over time in real-world scenarios, and it presents challenges in adapting to the diverse requests of consumers over time.
   
\noindent \textbf{Dynamic Popularity:} Dynamic popularity estimation captures the real-time changes in content popularity based on current demand \cite{Amadeo2022, Gui2020}. This approach involves routers considering various real-time factors, such as the total request count during a specific time interval, the current available cache capacity, the size of the requested content, etc, to calculate a dynamic popularity threshold value \cite{Ong2014}. This value determines the popularity status of content at a specific router or within a particular region. Unlike a static threshold, the dynamic threshold may vary between routers or differ across various regions within a network. The dynamic threshold value fluctuates over time in response to changing content trends, offering a more responsive and accurate estimation of content popularity. For example, a specific news article gains popularity during a certain period and subsequently declines in popularity. However, dynamic popularity estimation increases both communication and computational overhead, as routers need to periodically adjust the threshold value to track real-time changes.

\noindent \textit{4) \textbf{Probabilistic Caching:}} The probabilistic approach relies on the value of probability \cite{Li2024, Psaras2012} for content caching decisions instead of depending on deterministic rules. Routers can either use predetermined fixed probability values or random values within the interval of 0 to 1 for making caching decisions. Alternatively, routers can use a dynamic probabilistic approach, considering dynamic factors such as network traffic, cache capacity, and the distance between the consumer and provider to make content caching decisions.

\noindent \textit{5) \textbf{Network-Centric Caching:}} In this approach, content is stored within the network based on the topological structures and characteristics of routers, including the number of routers, their location, connectivity, link latency, etc. The goal is to cache content in close proximity to the consumer, reducing access time. By strategically placing content at important locations in the network, they enhance hit rates and reduce the load on the original content source. 

Based on graph-based metrics, these caching schemes evaluate the significance of routers within the network topology. Key metrics, such as degree centrality, closeness centrality, and betweenness centrality \cite{Naeem2024, Zheng2019}, play a crucial role in shaping caching decisions. These metrics help in placing and retrieving content strategically to improve network performance.

At the network level, caching schemes use either a consistent strategy for caching content at both the edge (closer to consumer) and core (closer to content source) of the network or implement a different approach based on specific requirements \cite{Amadeo2022}. In edge caching, content is stored at the network edge or in routers directly connected to consumers. On the other hand, core caching involves storing content in routers positioned at the core of the network, ensuring availability to a broader range of consumers across different regions. 

In order to enhance the efficiency of searching and caching operations, the network topology is viewed in smaller regions \cite{Yoshida2024} rather than the entire topology. This method enhances understanding of consumer demand relationships. One approach to implementing this involves a community-driven strategy, wherein the network topology is organized into distinct community structures or clusters \cite{Huang2019}. Another viable method is a neighborhood-based approach, establishing communication among neighbors within 1-hop, 2-hop, or N-hop distances \cite{Hu2020a, Chaudhary2023}.

\noindent \textit{6) \textbf{Content Granularity Level Caching:}} The level of information or amount of content kept in a cache is referred to as content granularity. This concept entails the identification of the specific content or information to be stored in the cache, aiming to enhance the efficiency of content retrieval. The granularity can vary, involving fine-grained approaches like chunk-level caching \cite{Cho2012}, where individual content chunks or blocks are cached, and coarse-grained strategies like object-level caching, where entire requested content objects or files are considered for caching \cite{Pavlou2013}. Another level of granularity, referred to as packet-level caching \cite{Arianfar2010, Thomas2015}, manages objects with finer granularity than object-level, which involves caching at the level of individual network packets. The decision regarding the extent of information to be stored in the cache is influenced by the requirements of the caching methodology.

\noindent \textit{7) \textbf{Freshness-Aware Caching:}} Content freshness is one of the most important factors in caching. These strategies consider the age of content. The content lifetime is defined with respect to the time when it is created by the content provider until its expiration time \cite{Zhang2023}. The content freshness parameter \cite{Quevedo2014, zhang2018iot} in the Data packet is set to indicate the lifetime of the content. During this time, routers can hold the content in their cache. Content freshness \cite{Alduayji2022, Amadeo2022} indicates how recent the content is at the time of caching. 

\noindent \textit{8) \textbf{Machine/Deep Learning-Based Caching:}} In recent years, ML/DL algorithms have been widely used in various areas of networking. For example, these algorithms are used to predict advanced cyber threats, detect unusual network behaviors, monitor networks in real time, optimize resource allocation, and many other applications. Along similar lines, researchers \cite{Hou2023intelligent, Hou2021, Liu2017b, Zhang2022} have used ML/DL techniques to make adaptive caching decisions in NDN to improve its performance. These adaptive algorithms can help make caching decisions based on real-time changes in user demand and network conditions. These algorithms help predict content demand and identify the best locations for caching to maximize benefits. Popular algorithms, such as Graph Neural Networks (GNN) \cite{Hou2021, Hou2022} and Reinforcement Learning (RL) \cite{Iqbal2023, Zhang2022}, are used to predict user demand and network conditions to select the most appropriate content and locations for caching. However, due to the complexity of these algorithms, conventional content caching methods remain popular choices. This is because the router can make caching decisions quickly while forwarding Data packets towards consumers, improving content availability and reducing response time.

\begin{figure*}[!h]
\centering
  \includegraphics{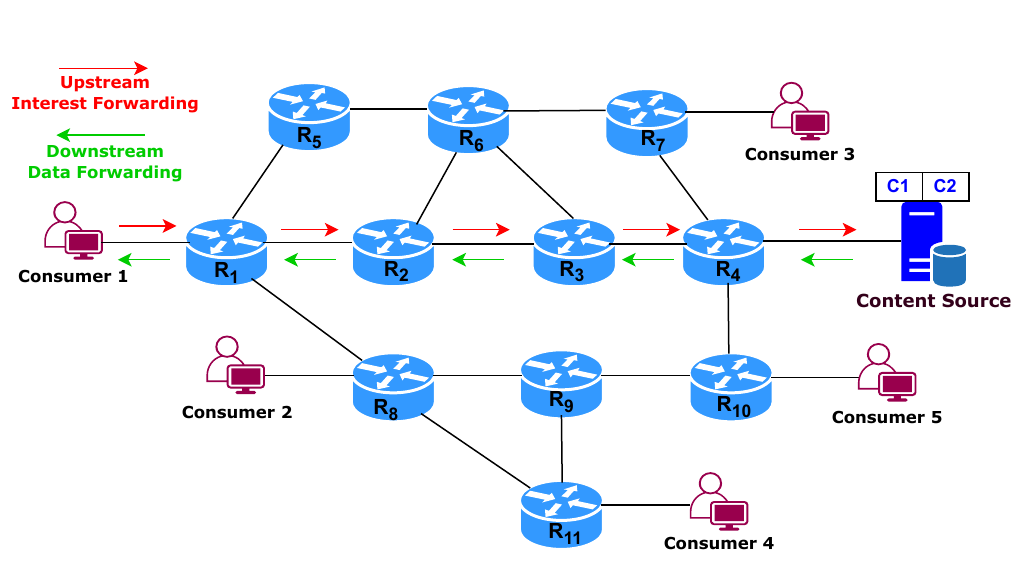}
  \caption{Reference Network Topology }
  \label{fig:RefArchitecture}
  \vspace{-4mm}
\end{figure*}

\begin{table*}[!t]
\centering
\caption{Summary of Various Caching Categories}
\label{tab:caching_definitions}
\resizebox{\linewidth}{!}{%
\begin{tabular}{|p{3.0cm}|p{2.0cm}|p{12.0cm}|p{2.5cm}|}
  \hline 
  \multicolumn{2}{|l|}{\textbf{Caching Category}}    & \textbf{Descriptions} & \textbf{Works} \\
  \hline
  \multicolumn{2}{|l|}{Centralized}      & The controller(s) or designated node(s) centrally manage content lookup and caching decisions. & \cite{Alhowaidi2021}, \cite{Jmal2017a}, \cite{Zhang2020a} \\
  \hline
  \multicolumn{2}{|l|}{ Decentralized}     & Distributed decision-making empowers multiple routers to perform content searching and caching decisions autonomously. &   \cite{Cho2012}, \cite{Ren2014} \\
  \hline
  \multicolumn{2}{|l|}{On-path}      & Content searching and caching are confined to routers situated on the communication path between the consumer and provider. In the downstream direction, content is either cached by all routers or selectively based on caching policies. &   \cite{Draxler2012}, \cite{Jacobson2009}, \cite{Laoutaris2004}, \cite{Psaras2012}\\
  \hline
  \multicolumn{2}{|l|}{Off-path}      & Provides flexibility in content exploration and caching. In the downstream direction, content can be cached anywhere in the network. &  \cite{Draxler2012}, \cite{Li2011}, \cite{Saino2013}, \cite{Yang2019a} \\
  \hline
   \multicolumn{2}{|l|}{Non-cooperative}      & Routers independently perform content lookup and caching operations without collaboration. &   \cite{Hubballi2024a}, \cite{Jacobson2009}, \cite{Zhang2017} \\
  \hline
  \multirow{2}{*}{Cooperative} & Implicit & Routers do not directly share content availability or any other information among themselves. &  \cite{Cho2012}, \cite{Laoutaris2006} \\
  \cline{2-4}
  & Explicit & Routers explicitly advertise their content availability and share other information. & \cite{Chaudhary2023}, \cite{Hu2020a}, \cite{Wang2013b}, \cite{chaudhary2025eencache}  \\
  \cline{2-4}
  \hline 
  \multicolumn{2}{|l|}{Popularity-Agnostic}      &  Treating all content equally, irrespective of the frequency of requests. &   \cite{Chaudhary2022} \\
  \hline
  \multirow{4}{*}{Popularity-Based} & Local & Considers consumer requests in a specific region, or each router individually tracks requests passing through it. &  \cite{Bernardini2013}, \cite{Ong2014}\\
  \cline{2-4}
       & Global & Considers content requests from all consumers across the entire network. &  \cite{Gui2020}, \cite{Hubballi2024b}, \cite{Li2012b}\\
  \cline{2-4}
       & Static Threshold & Utilizes a fixed predefined threshold value to determine which content to cache.  & \cite{Bernardini2013}, \cite{Ong2014}, \cite{Meng2023} \\
  \cline{2-4}
       & Dynamic Threshold & Adjusts the threshold value dynamically based on changing network conditions. & \cite{Hubballi2024a}, \cite{Ong2014} \\
  \hline 
   \multirow{3}{*}{Probabilistic-Based} & Fixed & Routers make caching decisions using a predetermined fixed probability value $p$. &  \cite{Laoutaris2006}, \cite{Tarnoi2014}\\
  \cline{2-4}
       & Random & Routers generate random number as probability value and this value of $p$ ranging between 0 to 1 is used for caching decisions. & \cite{Saino2014}, \cite{Zhang2015} \\
  \cline{2-4}
       & Dynamic &  Routers dynamically calculate the value of $p$ based on factors like popularity, cache size, network traffic, or distance. &  \cite{Psaras2012}, \cite{Tarnoi2019}, \cite{Wang2013c}\\
  \hline   
   \multirow{7}{*}{Network-Centric} & Edge Caching & Storing content at the network edge or the first upstream router to the consumer. &   \cite{Alduayji2023}, \cite{Saino2014}\\
  \cline{2-4}
       & Single Node & During content delivery, caching occurs at a single router in the network. &   \cite{Laoutaris2006}, \cite{Saino2013}\\
  \cline{2-4}
       & Multi-Node & During content delivery, caching occurs at multiple routers in the network. &   \cite{Jacobson2009}, \cite{Laoutaris2004}\\
  \cline{2-4}
       & Topology-Aware & Optimizes caching and retrieval by considering network topology attributes, including connectivity, link latency, and centrality metrics like degree, closeness, and betweenness. Routers with higher centrality values are prioritized for caching. & \cite{Chai2013}, \cite{Khandaker2021}, \cite{Zheng2019}\\
  \cline{2-4}
       & Location-Aware & Cache and retrieve content from specified routers in the network or domains. & \cite{Saino2013}, \cite{Sourlas2016} \\
  \cline{2-4}
       & Neighborhood-Aware & Caching and retrieval operations are influenced by the proximity of routers, ranging from 1-hop to N-hop neighbors. &  \cite{Chaudhary2023}, \cite{Hu2020a}, \cite{Mick2016}, \cite{Pal2021}\\
  \cline{2-4}
       & Community-Aware & Viewing network topology as multiple community/cluster structures based on node relationships for optimized content searching and caching decisions. & \cite{Huang2019}, \cite{Sourlas2016},  \cite{Xu2024}, \cite{Yan2017}, \cite{gui2025dynamic} \\
  \hline 
  \multirow{4}{*}{{Content Granularity}} & Object-Level & Routers cache entire objects or files. &  \cite{Pavlou2013} \\
  \cline{2-4}
   &  Packet-Level & Caching data at the level of individual network packets. &  \cite{Arianfar2010}, \cite{Jacobson2009}, \cite{Thomas2015}\\ 
  \cline{2-4}
   & Chunk-Level  & Routers cache large files by dividing them into chunks or blocks. & \cite{Cho2012}, \cite{Li2012a}, \cite{Li2011}\\ 
  \cline{2-4}
    & Network Coded & Chunks are encoded for caching in the router to improve data transmission efficiency. & \cite{Hu2020b}, \cite{Hu2020a}, \cite{Wu2013} \\
  \hline  
  \multicolumn{2}{|l|}{Freshness-Aware} & Content remains valid until the freshness period expires. &   \cite{Alduayji2023}, \cite{Amadeo2022} \\
  \hline
  \multicolumn{2}{|l|}{Machine \& Deep Learning-Based} & Using advanced learning methods like deep learning, reinforcement learning, and neural networks, routers can predict consumer needs, optimize content caching and retrieval, and adapt to real-time changes in user behavior. & \cite{Hou2023}, \cite{Iqbal2023}, \cite{Liu2017b}, \cite{gui2025dynamic}, \cite{awais2025iscc} \\
  \hline
\end{tabular}}
\end{table*}

\begin{table*}[!t]
  \centering
   \caption{Caching Scheme Classification}
    \label{tab:caching_classification}
  \resizebox{\linewidth}{!}{%
  \begin{tabular}
{|p{3.8cm}|>{\centering\arraybackslash}p{2cm}|>{\centering\arraybackslash}p{2.1cm}|>{\centering\arraybackslash}p{2cm}|>{\centering\arraybackslash}p{2cm}|>{\centering\arraybackslash}p{2cm}|>{\centering\arraybackslash}p{2cm}|>{\centering\arraybackslash}p{2cm}|>{\centering\arraybackslash}p{2cm}|}
    \hline
    \textbf{Caching Technique} & \textbf{Centralized} & \textbf{Decentralized} & \textbf{On-path} & \textbf{Off-path} & \multicolumn{2}{c|}{\textbf{Cooperative}} & \textbf{Non-Cooperative} & \textbf{Popularity} \\
    \cline{6-7}
    & & & & & \textbf{Implicit} & \textbf{Explicit} & & \\
    \hline
    LCE or CEE \cite{Jacobson2009} & $\times$ & \checkmark & \checkmark & $\times$ & $\times$ & $\times$ & \checkmark & $\times$\\
    \hline
    LCD \cite{Laoutaris2006} & $\times$ & \checkmark & \checkmark & $\times$ & \checkmark & $\times$ & $\times$ & $\times$\\
    \hline
    MCD \cite{Laoutaris2006} & $\times$ & \checkmark & \checkmark & $\times$ & \checkmark & $\times$ & $\times$ & $\times$\\
    \hline
    Prob(p) \cite{Tarnoi2014} & $\times$ & \checkmark & \checkmark & $\times$ & $\times$ & $\times$ & \checkmark & $\times$\\
    \hline
    Prob-PD \cite{Ioannou2014} & $\times$ & \checkmark & \checkmark & $\times$ & \checkmark & $\times$ & $\times$ & \checkmark\\
    \hline
    ProbCache \cite{Psaras2012} & $\times$ & \checkmark & \checkmark & $\times$ & \checkmark & $\times$ & $\times$ & $\times$\\
    \hline
    MPC \cite{Bernardini2013} & $\times$ & \checkmark & \checkmark & \checkmark & $\times$ & \checkmark & $\times$ & \checkmark\\
    \hline
    FGPC/D-FGPC \cite{Ong2014} & $\times$ & \checkmark & \checkmark & $\times$ & $\times$ & $\times$ & \checkmark & \checkmark\\
    \hline
    Intra-AS \cite{Wang2013b} & $\times$ & \checkmark & \checkmark & \checkmark & $\times$ & \checkmark & $\times$ & $\times$\\
    \hline
    O2CEMF \cite{Hu2020a} & $\times$ & \checkmark & \checkmark & \checkmark & $\times$ & \checkmark & $\times$ & $\times$\\
    \hline
    NACID \cite{Pal2021} & $\times$ & \checkmark & \checkmark & \checkmark & $\times$ & \checkmark & $\times$ & $\checkmark$\\
    \hline
    eNCache \cite{Chaudhary2023} & $\times$ & \checkmark & \checkmark & \checkmark & $\times$ & \checkmark & $\times$ & $\times$\\
    \hline
    Hash-Routing (HR) \cite{Saino2013} & $\times$ & \checkmark & \checkmark & \checkmark & $\times$ & $\times$ & \checkmark & $\times$\\
    \hline
    CPHR \cite{Wang2015} & $\times$ & \checkmark & \checkmark & \checkmark & \checkmark & $\times$ & $\times$ & \checkmark \\
    \hline
    OpenCache \cite{Yang2019a} & $\times$ & \checkmark & \checkmark & \checkmark & $\times$ & \checkmark & $\times$ & \checkmark \\
    \hline
    OFAM-CCN \cite{Jmal2017a} & \checkmark & $\times$ & \checkmark & \checkmark & $\times$ & \checkmark & $\times$ & \checkmark \\
    \hline
    Compound Popularity \cite{Gui2020} & $\times$ & \checkmark & \checkmark &$\times$& $\times$ & \checkmark & $\times$ & \checkmark \\
    \hline
    SDN-Based Caching \cite{Zhang2020a} & \checkmark & $\times$ & \checkmark & \checkmark & $\times$ & \checkmark & $\times$ & \checkmark \\
    \hline
    CCS/CES \cite{Amadeo2022} & $\times$ & \checkmark & \checkmark &$\times$& \checkmark & $\times$ & $\times$ & \checkmark \\
    \hline
     SDCC \cite{Sharif2022} & \checkmark & \checkmark & \checkmark &\checkmark & $\times$ & \checkmark & $\times$ & \checkmark \\
    \hline
    PF-EdgeCache \cite{Alduayji2023} & $\times$ & \checkmark & \checkmark &$\times$& $\times$ & $\times$ & \checkmark & \checkmark \\
    \hline
    PTF \cite{Wu2023} & $\times$ & \checkmark & \checkmark &$\times$& \checkmark& $\times$ & $\times$ & \checkmark \\
    \hline
    SoftCaching \cite{Rafique2023} & \checkmark & $\times$ & \checkmark &\checkmark & $\times$& \checkmark & $\times$ & $\times$ \\
    \hline
    PePC \cite{Hubballi2024a} & $\times$ & \checkmark & \checkmark &$\times$& $\times$& $\times$ & \checkmark & \checkmark \\
    \hline
    EABC \cite{He2024} & $\times$ & \checkmark & \checkmark & $\times$ & \checkmark & $\times$ & $\times$ & $\times$ \\
    \hline
    CRUS \cite{Li2024} & $\times$ & \checkmark & \checkmark & $\times$ & \checkmark & $\times$ & $\times$ & \checkmark \\
    \hline
    
\end{tabular}}
\end{table*}

\subsection{Overview of Existing Caching Strategies}
In this subsection, we describe several important caching techniques designed by the ICN community. These caching methods fall under two broad categories: the one that does not take popularity into account and the second ones that take popularity into account. We elaborate the working of different approaches of two categories by taking reference to a small topology shown in Fig.~\ref{fig:RefArchitecture}. Table~\ref{tab:caching_classification} presents the summary of existing caching techniques under different categories. A tick mark (\checkmark) in the table indicates that the caching technique falls under that category.

\noindent \textbf{1) Popularity-Agnostic Caching:} These caching methods do not consider the frequency of requests while making caching decisions.\\
\noindent {\textbf{Leave Copy Everywhere (LCE) \cite{Jacobson2009}}:} LCE is alternatively referred to as Cache Everything Everywhere (CEE). This is a non-cooperative on-path caching approach, where routers positioned between the content consumer and provider (router/original source) cache all content passing through them in the downstream direction. LCE involves caching the same content in all routers along the transmission path. In Fig.~\ref{fig:RefArchitecture}, Consumer1 initiates a request (C1), which is served by the source. The request is routed via the shortest path through routers $R_1$, $R_2$, $R_3$, and $R_4$ to reach the provider. In response, content C1 is cached at $R_4$, $R_3$, $R_2$, and $R_1$, then served to Consumer1. Implementing LCE is straightforward, as routers do not consider neighbors in caching decisions. However, caching the same content at multiple routers increases redundancy and eviction rates, impacting the overall performance.

\noindent {\textbf{Leave Copy Edge (LCEdge) \cite{Saino2014}}:}
LCEdge is a single-node on-path caching approach that caches content at the last router in the downstream direction, typically one hop away from the consumer, especially when the directly connected router to the consumer is cache-enabled. In Fig.~\ref{fig:RefArchitecture}, if Consumer1 requests content C1, it is provided by the source node. In the downstream direction, the content is cached only at router $R_1$, while other routers on the path forward the packet to the next hop without caching. LCEdge is designed to minimize content access time by keeping content closer to the consumer. However, it underutilizes other routers on the delivery path that are also capable of holding content.

\noindent {\textbf{Leave Copy Down (LCD) \cite{Laoutaris2006}}:} LCD is another popular on-path caching technique specifically designed to address the content redundancy challenges posed by LCE. Unlike LCE, LCD strategically caches content at only one router along the delivery path. In the case of LCD, the content is cached one hop down from the content provider in the downstream direction, while other routers along the path forward the packet without caching it. The effectiveness of LCD relies on implicit cooperation with the next hop router, facilitated by the inclusion of a flag bit in the packet. This flag informs routers on the forwarding path that the upper-level router has already cached the content. LCD brings frequently requested content closer to the consumer upon subsequent hits. It achieves this without the need to monitor content popularity through the maintenance of additional tables. Instead, it intuitively brings popular content closer to consumers with each subsequent hit, effectively increasing the hit ratio and minimizing content access time. To illustrate this concept, let's refer to Fig.~\ref{fig:RefArchitecture}. In the initial round, Consumer1 requests content C1, which is initially available at the source node. As it travels downstream, C1 is cached one hop down at $R_4$. In the second round, Consumer1 again requests C1, which is now provided by $R_4$ and cached at $R_3$, and this is repeated until the content is brought to $R_1$ through successive requests.  However, it becomes apparent that the same content is cached at all routers along the transmission path, leading to content redundancy similar to the LCE. Another concern with LCD is that if consumers request different content each time, for example, in a round-robin fashion, caching and eviction operations only occur at a specific router (e.g., $R_4$). This can result in a reduced cache hit ratio and increased content access time.

\noindent {\textbf{Move Copy Down (MCD) \cite{Laoutaris2006}}:} MCD is another strategy that caches content at a single router along the delivery path. MCD reduces redundancy compared to LCD by deleting content from the provider node after caching it one hop down the router in the downstream direction, except when the content is served by the original source. 

\noindent {\textbf{Energy-aware Approximate Betweenness Centrality (EABC) \cite{He2024}}:} EABC is a caching technique designed for NDN-based IoT that considers router centrality and energy availability when making caching decisions. In request forwarding, the EABC strategy adds a \texttt{Betw} field to the Interest packet, which helps to select the router with a higher betweenness centrality along the path. When an Interest packet passes through a router, each router along the path compares its centrality value with the value present in the \texttt{Betw} field. If the router's centrality value is higher than the \texttt{Betw} value, it updates the \texttt{Betw} field; otherwise, it forwards the Interest packet to the next hop without updating. In content forwarding, the EABC strategy adds two fields, \texttt{Betw} and \texttt{EnergyFlag}, to the Data packet to make caching decisions based on the router's centrality and energy level. The \texttt{Betw} field carries the centrality value of the router, and the \texttt{EnergyFlag} is used to determine whether the packet is new (sufficient energy) or old (insufficient energy). \texttt{EnergyFlag} value is set to 0 at the beginning, indicating that the packet is new. On subsequent hops, the value changes depending on the energy level of the router. The value of the \texttt{EnergyFlag} is set to 0 if the router's energy is higher than a predefined threshold; otherwise, it is set to 1, indicating that the router will not cache the content due to insufficient energy. At each hop, if the router's energy is lower than the predefined threshold, the value of the \texttt{EnergyFlag} is incremented by 1. When the \texttt{EnergyFlag} value exceeds 2, the content will be cached regardless of any conditions to improve the performance of caching.

\noindent \textbf{Probabilistic-Caching:}
Works in \cite{Psaras2012, Tarnoi2014} use the probabilistic algorithm where router $R$ makes caching decisions for content $C$ based on probability $p \in [0,1]$.

\textsl{Prob(p)~\cite{Laoutaris2006, Tarnoi2014}:} Prob(p) enhances the conventional approach (caching everything along all routers in the downstream direction) by selectively caching content with a fixed probability value $p$. This probability-based caching significantly decreases content redundancy. In the downstream path, each router independently generates a random number between 0 and 1. If this generated number is less than $p$, the router caches the content; otherwise, it refrains from caching. Like LCE, Prob(p) operates in a non-cooperative manner, with each on-path router autonomously making caching decisions. This decentralized decision-making process avoids communication overhead. Notably, Prob(p) significantly reduces content redundancy and replacement rates compared to LCE. The effectiveness of Prob(p) depends on the predefined value of $p$; when $p$ reaches one, Prob(p) becomes equivalent to LCE. Because of this adaptability, Prob(p) can exhibit varied performance characteristics depending on the given probability threshold.

\textsl{ProbCache~\cite{Psaras2012}:} ProbCache is another popular on-path probabilistic caching method that caches content on the most efficient router based on the characteristics of the content delivery path. ProbCache brings content closer to the consumer based on the estimated probability. In order to monitor the number of hops the request and content packets travel, the ProbCache introduces two additional fields in the ICN packet structure: Time Since Inception (TSI) and Time Since Birth (TSB). In the upstream direction, the TSI value increases by one at each crossing router, while in the return path, the TSI value remains fixed, reflecting the distance between the consumer and provider. Simultaneously, in the downstream direction, the TSB value increments at each crossing router. By utilizing TSI and TSB values, the ProbCache calculates \texttt{TimesIn}, representing the number of times the path can accommodate caching for a given packet. It also considers the cache weight factor to ensure a fair distribution of resources. Each router makes caching decisions based on the estimated probability, determined by the product of \texttt{TimesIn} and the cache weight factors.
To enhance the performance of the ProbCache \cite{Psaras2012} strategy, an extension called ProbCache$+$ \cite{Psaras2014} was introduced. ProbCache$+$ improves performance by modifying the cache weight factor.

ProbCache uses implicit cooperation techniques wherein routers do not externally advertise their cache status to neighboring routers. This approach reduces communication overhead. Caching content near the consumer along the on-path route increases the hit rate. However, a limitation of ProbCache is its lack of consideration for the popularity level of content in the network, making it unable to distinguish between popular and unpopular content. This absence of content popularity distinction may potentially impact its overall performance.

\noindent \textbf{Neighborhood-Based Cooperative Caching:} The works presented in \cite{Chaudhary2022, Chaudhary2023, Hu2020a, Wang2013b} strategically explore nearby off-path routers to retrieve available content while forwarding requests towards the provider.

Hu \emph{et~al.}~\cite{Hu2020a} proposed an on-demand off-path cache exploration technique termed O2CEMF for the network coding enabled NDN. Network coding divides large files into multiple chunks, which are then encoded before being sent. This enables multi-path forwarding of several chunks simultaneously. O2CEMF discovers encoded blocks present in off-path routers while forwarding request packets to the original content source. It uses a multipath packet transmission approach in which Interest packets are simultaneously forwarded to both upstream on-path and off-path routers. O2CEMF operates in two phases: exploration and exploitation. During the exploration phase, each router situated on the request forwarding path sends O2CEMF Interest packets to all its neighboring routers. The off-path routers respond with either encoded block information from their caches or a NACK packet if there are insufficient blocks available for the requested generation. Using this information, the router constructs the cache trails table, which proves beneficial in the exploitation phase. In the exploitation phase, routers send Interests based on matching entries in the cache trails. However, due to the high volatility of cached content, O2CEMF may introduce signaling overhead due to the flooding of Interest packets.

NCache \cite{Chaudhary2022}, eNCache \cite{Chaudhary2023} describe efficient cooperative content searching and caching approaches for NDN. These methods aim to explore nearby off-path neighbors for the requested content while forwarding the request to the content source. While NCache is limited to 1-hop off-path exploration, whereas eNCache extends to K-hop off-path neighborhood lookup in an optimized way. eNCache sends requests simultaneously to both on-path and off-path neighbors to retrieve the content. If the content is not available in any  off-path and on-path routers, eNCache directly retrieves it from the original content source. eNCache uses different Interest packet structures to facilitate content search among on-path and off-path neighbors. In the off-path Interest packet structure, eNCache sets the hop limit field to specify the depth to which off-path neighbors are queried. Additionally, it introduces the \texttt{NoPITEntry} field to ensure that off-path routers do not forward and create entries for their neighbors.
In order to enhance content searching within the neighborhood, eNCache introduced the \texttt{VisitedNodes} field to the on-path Interest packet structure. This field tracks all the routers visited up to the previous hop. As each on-path router forwards a request, it adds IDs of all the visited routers to this field and transmits this information to the upstream router. 
This ensures that the upstream router does not forward the packet to the routers specified in the \texttt{VisitedNodes}, thus avoiding the generation of duplicate packets at the router and reducing the wastage of network resources.

Wang \emph{et~al.}~\cite{Wang2013b} proposed a similar off-path searching method up to a 1-hop neighborhood to retrieve requested content before forwarding it to the original source. However, this is limited to within the Intra-Autonomous System (Intra-As), enabling routers to probe within this Intra-AS framework. Neighboring routers collaborate to satisfy each other's requests while also reducing content redundancy in the neighborhood. Routers periodically share their cache table information with directly connected routers using the Bloom filter data structure, effectively minimizing the overhead associated with information sharing.

Badshah \emph{et~al.}~\cite{Badshah2019} proposed a centralized caching method for SDN-based ICN architecture. In this solution, multiple cache servers are deployed in the topology instead of a single centralized cache server \cite{Kim2017} to distribute the load and enhance system efficiency. The proposed approach begins by calculating the required number of cache servers based on the size of the network topology. Subsequently, it deploys these cache servers strategically, considering multiple metrics such as path stretch, closeness centrality, and betweenness centrality. This strategic placement aims to reduce traffic and content access time by situating cache servers closer to routers. 

Once the cache servers are appropriately located, the network controller is responsible for installing flow rules in every switch to direct requests to the nearest cache server. The controller, acting as a global entity, keeps track of cache server and content provider locations in the network. Upon receiving a request, if the requested content is found in the cache server, it is returned to the requesting node. However, if the content is not found in the cache, the request is then forwarded to the controller. The controller then determines the path between the cache server and the appropriate content provider for handling the request and content routing. In the response path, the content provider first sends the content to the cache server for caching purposes. Subsequently, the cache server forwards the content to the requesting node. 

Rafique \emph{et~al.}~\cite{Rafique2023} introduced a centralized approach known as SoftCaching for IoT networks, which integrates SDN with ICN to efficiently perform the routing and caching operations. With the help of a controller, content is cached at selected routers, which are considered the best for caching. SoftCaching focuses on selecting the best caching nodes based on network traffic analysis. The controller, possessing a global view of the network, plays a crucial role in identifying caching nodes, installing flow rules in switches for routing requests to the nearest caching node, and overseeing all ICN-related operations. In the SoftCaching framework, the choice of best nodes for caching is determined through an analysis of network traffic, employing methods such as Singular Value Decomposition (SVD) and QR Factorization. SoftCaching aims to improve the effectiveness of caching by strategically choosing caching nodes through a thorough assessment of network traffic patterns.

Authors of \cite{Lee2025} designed a caching technique called Heartbeat for NDN to access content available in the caches of off-path routers. Heartbeat uses a proactive advertisement approach that enables effective utilization of off-path caches. However, it employs a lightweight mechanism to share content availability information among neighboring nodes. Instead of sending entire catalogs periodically, Heartbeat sends virtual interests to neighboring nodes whenever content is inserted or relocated in a cache. Neighboring nodes use this virtual interest to create temporary FIB entries with carefully chosen lifetimes. This enables user requests to be redirected to nearby off-path caches without causing flooding. The simulation results demonstrate that Heartbeat improves cache utilization while reducing bandwidth consumption and overhead.

\noindent {\textbf{Hash-Based Techniques:}} Hash-based methods utilize a hash function to map requested content identifiers to values, enabling the identification of routers within the network. This location-aware method allows routers to determine where to forward Interest packets and where to cache the content in the network. Various hash-based techniques \cite{Li2016, Saha2015, Saino2013, Sato2019, Sourlas2016, Wang2015} have been designed to locate content in the network. These techniques fall under the category of off-path caching approaches, as content can be retrieved and cached from anywhere in the network, not necessarily always on the transmission path.

Another method by Saino \emph{et~al.}~\cite{Saino2013} assigns every router in the domain or topology a unique ID between 1 and N, where N represents the total number of nodes in the topology. For every unique content, one router is designated to cache the content in its CS based on the hash value of the request. When an edge router receives a request for content, it first calculates the hash value of the requested content by passing the request identifier to a hash function. This hash value is used for locating the designated cache node. Subsequently, the edge router routes the request packet to the designated cache node via the shortest path. If the content is available at the designated cache node, it replies; otherwise, the request packet is forwarded all the way to the source. They evaluated five hash-based schemes: Symmetric, Asymmetric, Multicast, Hybrid Symmetric-Multicast, and Asymmetric-Multicast. While these techniques use the same routing mechanism for forwarding content requests, they differ in the way they forward content packets.  The Symmetric technique follows the same path for both request and content forwarding. In this approach, the content is first forwarded to the designated node for caching and then sent to the requested edge router. In the Asymmetric scheme, the content packet follows the shortest path between the provider and the edge router. If the designated caching node is located on the shortest path, it caches the content; otherwise, this technique ignores caching. The Multicast approach involves forwarding one copy of the content to the designated node for caching and another copy to the requested edge router via the shortest path. Hybrid techniques combine the best aspects of these methods to enhance both routing and caching performance.

Hash-based schemes make use of the content available at off-path routers to improve the cache hit ratio and reduce cache redundancy by caching a single copy of the content in a network \cite{Saino2013}. However, these schemes suffer from increased content access time in larger network domains/topologies due to the designated node potentially being distant from the edge router. To overcome these challenges, Sourlas \emph{et~al.}~\cite{Sourlas2016} proposed a hash-based routing approach that incorporates domain clustering methods. This approach reduces content access time caused by hash-based routing schemes \cite{Saino2013} while maintaining a higher cache hit ratio. In this technique, larger domains are divided into several clusters or groups, and a modulo hash function is used to map content requests to the designated cache node within the respective cluster. They use well-known clustering algorithms such as k-split and k-medoids to partition the large network topologies.

Wang \emph{et~al.}~\cite{Wang2015} presented a cooperative hash-based routing and caching scheme named CPHR with the objective of maximizing the cache hit ratio by eliminating content redundancy within the network. CPHR partitions the content space based on hash values and assigns each partition to the cache-enabled routers in the network. The process of forwarding Interest and content packets in CPHR follows a symmetric scheme, similar to the approach outlined in \cite{Saino2013}. To enhance request forwarding, CPHR augments the FIB table by introducing an extra field called \texttt{EgressRouter}. This field guides the routing of Interests associated with a specific name prefix, directing them to the designated egress router before exiting the domain when forwarded towards the source node. 
Furthermore, each ingress router maintains a partition table to identify the partition of the received request and the corresponding assigned cache router. CPHR also adds two additional fields to the NDN Interest packet: \texttt{CacheName} indicating the name of the assigned cache, and \texttt{EgressName} denoting the name of the egress router. The assigned cache router updates the \texttt{EgressName} field set by the ingress router (edge router) to guide the Interest towards the original content provider node. \\

\noindent {\textbf{ 2) Popularity-Based Caching:}} 
These caching schemes incorporate content popularity (demand of content) as one of the key factors while making caching decisions. 

\noindent\textbf{Most Popular Content (MPC)~\cite{Bernardini2013}:} MPC proposes an approach to cache only popular content to enhance the cache hit ratio. Each router maintains a Popularity Table (PT) recording content names and their access counts. If an entry already exists, the counter is incremented. MPC relies on a predefined threshold to determine local content popularity. If the request count surpasses this threshold, the content is considered popular and it is cached. The router then suggests to its neighbors to cache this popular content. The caching decision in the neighbor depends on the local policy (whether to cache or not). After making suggestions to neighbors, the router resets the popularity count to prevent unnecessary flooding of suggestions for the same content. \\
In Fig.~\ref{fig:RefArchitecture}, Consumer1 requests C1, which is available at the source node. For instance, at router $R_4$, the count of content C1 is 5, surpassing the set threshold of 5, categorizing C1 as popular and prompting its caching at $R_4$. Subsequently, $R_4$ suggests its neighbors, $R_3$, $R_7$, and $R_{10}$, to cache C1. This approach enables MPC to cache content at routers beyond the delivery path, like $R_7$ and $R_{10}$. Although MPC minimizes overhead by limiting popularity measures locally, caching only popular content may not optimally utilize the available cache space. Additionally, MPC might introduce overhead when disseminating caching suggestions to neighboring routers.\\
\textbf{Fine-Grained Popularity-Based Caching (FGPC)~\cite{Ong2014}:} Depending on the available cache space, FGPC strategically caches both unpopular and popular content to optimize cache utilization and hit rate. FGPC caches all passing content when space is available, and when the cache is full, it removes unpopular content while caching popular content using the  Least Recently Used (LRU) replacement policy. To monitor the content frequency, each router maintains a Popularity Table (PT), similar to the MPC \cite{Bernardini2013}. Content is considered locally popular to cache when the local count value of requested content reaches a predefined threshold. As content popularity changes over time, a static threshold may diminish performance. To address this, FGPC has been extended to D-FGPC \cite{Ong2014}, which caches content based on a dynamic threshold. This dynamic threshold adjusts depending on the available cache space, content size, and the request count of content in the PT at time T. However, the dynamic threshold introduces increased overhead at the routers.\\
\textbf{Popularity-based Predictive Caching (PePC)~\cite{Hubballi2024a}:} 
In PePC, routers along the delivery path dynamically adjust caching decisions based on the current occupancy of the cache. It operates as a non-cooperative technique, where each router independently makes caching decisions and maintains a popularity table to track content requests. The caching decision model in PePC draws inspiration from the Random Early Detection (RED) \cite{Floyd1993} algorithm. PePC utilizes minimum and maximum thresholds to determine caching behavior: if cache occupancy is below the minimum threshold, all content is cached to maximize space utilization; if it falls between the minimum and maximum thresholds, content is cached based on predicted values and reaching the maximum threshold triggers caching only of popular content due to critical cache space. While PePC avoids communication overhead, its non-cooperative nature may lead to content redundancy, as each router might cache the same set of contents.

\noindent \textbf{Probabilistic-Caching:} Ioannou \emph{et~al.} \cite{Ioannou2014} introduced the on-path probabilistic caching scheme termed Prob-PD to enhance the performance of on-path caching. While the content is delivered in the reverse path, on-path routers dynamically calculate probabilities based on two key factors: the popularity of the requested content at the router, making the caching decision, and the distance ratio of the router from the content source. 
Another work \cite{Li2024} proposed a probabilistic caching technique called the Caching-Resource Utilization-Based Strategy (CRUS). In this approach, the on-path router with the lowest resource utilization is selected for content placement. The cache resource utilization in CRUS is estimated based on the number of requests served by a router, $R_i$, relative to its cache occupancy. Content is cached in the on-path router based on probability. If the content is served by the source, the probability of caching is 1. However, if the content is served from the cache of a router, the caching probability is determined by comparing the number of hops between the selected router for content placement and the provider node to the number of hops between the provider and the consumer. The closer the selected router is to the consumer, the higher the probability of caching the content.

\noindent \textbf{Neighborhood-Based Cooperative Caching:}
Work~\cite{Yang2019a} introduced OpenCache, a lightweight collaborative caching method to retrieve popular content from nearby locations efficiently. The main idea behind this strategy is the exchange of hierarchical name prefixes among routers. By sharing these prefixes, routers make popular content accessible to other routers within the network. Neighboring routers then maintain a record of these shared prefixes along with their corresponding interfaces, directing subsequent requests to the nearby available caches. OpenCache achieves lightweight off-path cache collaboration through the extension of the original FIB with open cache entries. These entries encompass both open data prefixes and associated cache interfaces. 

Another work \cite{Reshadinezhad2023} proposed a neighborhood collaboration scheme to achieve efficient content distribution by forming local clusters among adjacent routers (1-hop neighbors). Each router shares its cache information with neighboring routers whenever new content is cached or replaced  by sending a \texttt{Content-Ad} message. Each router maintains a Neighbor Cache Table (NCT) to keep track of content availability in adjacent routers and cost of fetching that content. The NCT helps construct the FIB table to retrieve content from nearby routers located outside the transmission path (on-path). While forwarding a request towards the content provider, the router first checks its cache; if the content is not available, it checks for the content in any adjacent router. If the content is not available in any neighboring routers, the request is routed towards the provider. In the downstream direction, each router decides whether to cache content based on the state (True/False) of the \texttt{NeighborCached} flag, which is added to the Data packet and a threshold value. This flag helps avoid caching the same content in adjacent routers. The caching threshold is computed by considering several key factors, such as router connectivity, content popularity, hop count, and cache replacement rate, which helps in caching the popular content at important routers.

Yoshida \emph{et~al.} \cite{Yoshida2024} proposed PopDCN, in which the network topology is divided  into multiple clusters of the same size to begin with.  Each router is assigned a unique ID in the cluster for helping in content caching. Subsequently these cluster sizes are varied (increased or decreased) based on the popularity of contents stored in these nodes.  Every piece of content chunk is mapped to a specific router within the cluster using a hash function. This ensures no content redundancy within the cluster, as each content chunk is cached at a single location determined by the hash function. PopDCN first tries to retrieve content from the originating cluster. If this fails, the request is forwarded to the nearest router in another cluster. PopDCN uses a cache advertisement approach to efficiently retrieve content, where each router updates its nearest neighbor about its cache state whenever there is a change. In this work, the authors used a grid topology for simulation, where clusters can be easily divided into equal sizes. However, in a real network, the main challenge lies in forming clusters while keeping their sizes balanced.

Gui and Chen \cite{Gui2020} developed an on-path content placement scheme that incorporates node and content popularity. This strategy considers both local and global content request patterns. Local popular content is cached at the edge router to fulfill requests from local region consumers, while globally popular content is cached at routers based on the estimated cache benefit. The cache benefit is determined by considering requests from all consumers and the node's popularity. Global popularity information is gathered by establishing communication among the edge routers within the topology. Each edge router maintains an Interest Statistical Table (IST) to track local content requests, and at the end of each cycle, edge routers exchange ISTs with each other. Additionally, all routers maintain an Information Record Table (IRT) to track the entry face and hop count of Interest packets along the route. In the downstream direction, content is cached in routers based on the \texttt{Cache nodes} field of the Data packet. This field contains the ID of the router where the content will be cached in the downstream direction. If it carries more than one router, caching occurs as follows: for globally popular content (beyond the dynamic threshold), it will be cached at all routers; for potentially globally popular content, it will only be cached at the first router in the returning path. If the content is globally popular and if there is no router in the \texttt{Cache nodes} field, it is cached at the downstream router of the provider node.

The communication overhead in exchanging global popularity information among edge routers is contingent on the number of cycles. A higher number of cycles enables a more accurate collection of request patterns but also introduces additional communication overhead. Conversely, fewer cycles reduce communication overhead but may impact performance as routers exchange data over longer intervals. 

In \cite{Jmal2017a} a centralized caching technique OFAM-CCN is presented. This uses the OpenFlow architecture to manage content caching. OFAM-CCN implements a centralized entity called the Cache Management (CM) table, responsible for caching popular contents within the domain. CM keeps track of the popularity table to determine which content is popular among consumers within the domain. Each time the OpenFlow controller receives a request, it forwards it to the CM. Upon receiving the request, the CM updates the popularity table and checks its cache; if the requested content is present in its cache, it forwards both the popularity information and the content chunks to the OpenFlow controller. In a similar spirit, another study \cite{Zhang2020a} proposed a centralized caching decision for ICN leveraging the SDN framework.  The SDN framework provides a global view of the network, which prioritizes caching the popular video in the network while considering the perspective of global consumers. The SDN controller is responsible for making caching decisions, routing requests to the best node in the network, and tracking both global and local content popularity by receiving statistics from all nodes periodically. The authors of \cite{Sharif2022} proposed a cache placement technique termed SDCC, which uses an SDN controller to cache popular content over time in ICN-IoT networks. They proposed two methods for managing the cache: centralized and decentralized. A centralized (single-controller) approach manages the global view of the network by placing the control logic at a central controller, while a decentralized (distributed-controller) approach reduces overhead and addresses the single point of failure. In the decentralized approach, the network is divided into multiple clusters, each managed by its own controller. Works outlined in \cite{Amadeo2022, Alduayji2023, Wu2023} consider both the popularity and freshness of content when making caching decisions.

Amadeo \emph{et~al.}~\cite{Amadeo2022} implement distinct caching strategies for core and edge routers. Core network routers independently decide on caching based on content popularity and freshness, reducing overhead. As core routers handle a substantial number of requests, caching popular and long-lasting content enhances the cache hit ratio. Popularity and freshness are determined by a dynamic threshold periodically estimated in time interval \texttt{T}. If both values reach the threshold, content is always cached; if below, it's never cached. For popular but less fresh content, the caching decision is based on probability. Edge routers promote content diversity through limited cooperation. Routers set the \texttt{CACHED} flag to true after caching content to prevent duplication by other on-path routers.

Alduayji \emph{et~al.}~\cite{Alduayji2023} introduce a caching scheme to bring popular content closer to the network's edge, facilitating efficient content retrieval with fewer hops. Each router manages a compact popularity table to monitor request frequencies, alleviating the computational load. Once  this table is full, routers use the Least Frequently Used (LFU) policy for content eviction, prioritizing the removal of less accessed entries to accommodate new requests. Content is cached if the popularity count reaches the dynamic threshold; otherwise, caching is avoided. When router capacity is reached, the least popular content is replaced with more popular content, ensuring a fresh cache. This strategy results in an increase in content popularity counts near consumers and a decrease in counts for routers away from the consumers.

Wu \emph{et~al.}~\cite{Wu2023} introduced a caching technique wherein routers make content caching decisions based on dynamically estimated cache benefit values. The priority for caching is determined by a higher cache benefit, which is calculated by considering factors such as content popularity, freshness, and the hop count between the consumer and the router.

Works~\cite{Wang2022, Amadeo2021b, Dhara2023, Liao2024} proposed caching techniques for vehicular NDN (VNDN), considering content request preferences and node mobility to enhance content distribution and cache diversity in high-mobility environments. Wang \emph{et~al.}~\cite{Wang2022} suggested a Popularity-Incentive Caching Strategy (PICS) for VNDN to cache content in vehicle nodes, thereby reducing content delivery time and the load on the base station. In PICS, the base station acts as a leader, announcing the price of content chunks to other nodes in the network for making caching decisions. Based on the announced price, vehicles make caching decisions by considering additional factors such as location information and content popularity. To enhance content diversity, Dhara \emph{et~al.}~\cite{Dhara2023} proposed a caching method, POPS-cache, where popularity estimation is done based on content categories (e.g., sports, social media, education, technology) within each region rather than the popularity of individual chunks. Amadeo \emph{et~al.}~\cite{Amadeo2021b} introduced DANTE, a caching mechanism for VNDN that aims to improve content diversity among neighboring vehicles. In DANTE, vehicles make caching decisions based on content lifetime,  popularity, and  availability of content in the neighboring vehicles. Liao \emph{et~al.}~\cite{Liao2024} designed EADC, a dynamic caching approach for VNDN, to efficiently handle frequently changing content demand in vehicular networks. To determine the probability of caching the content in the vehicle, EADC takes into account multiple factors, such as vehicle characteristics (social attributes and content popularity), motion centrality (degree and betweenness centrality), and transmission cost (transmission delay, transmission difficulty, and cache redundancy).

\section{Discussion}
\label{futuredirections}
In this section, we discuss  the performance metrics commonly used by researchers to evaluate caching techniques in ICN/NDN.  Taking reference to these performance metrics, we discuss the limitation of current works and provide future directions for further work in this direction.

\subsection{Evaluation Metrics} \label{metrics}
There are a number of performance metrics used for evaluating different caching techniques. These metrics assess effectiveness of the caching techniques. Following are the major performance metrics. 

 \noindent \textit{1) Cache Hit Ratio:} It measures the total content served by the routers in the network in relation to the overall consumer requests. A higher cache hit ratio indicates a decreased load on the content server as a larger proportion of content is successfully retrieved from the local caches.\\
    \noindent \textit{2) Cache Diversity:} This metric reflects the overall proportion of unique content present across all routers in the network. A higher cache diversity indicates lower content redundancy and improved cache hit ratio, as routers can serve a substantial portion of content requests.\\
    \noindent \textit{3) Latency (Roundtrip Time):} It assesses the total time taken for the request to reach the content provider and the subsequent time for the consumer to receive the corresponding data. Lower retrieval times indicate a more efficient content transfer from the nearby provider to the consumer.\\
    \noindent \textit{4) Hop Count:} The metric assesses the number of hops a content request takes to retrieve the desired content. Lower hop counts indicate that content is retrieved from the nearest provider, ensuring an efficient retrieval path.\\
    \noindent \textit{5) Hit Distance:} Hit distance measures the distance between the content provider (router/original source) and the consumer. The shorter the hit distance, the closer the content provider is.\\
    \noindent \textit{6) Hop Reduction Ratio:} It measures the reduction in the number of hops required to retrieve content compared to fetching it directly from its original source. This reduction is attributed to the availability of the content closer to the requester in the network.\\
   \noindent \textit{7) Content Delivery Ratio:} It evaluates the ratio of content requests successfully delivered (Satisfied Interest) to the total number of requests made. A higher content delivery ratio indicates lower request loss during transmission.\\
\noindent \textit{8) Cache Replacement:} Cache replacement (eviction) measures how often content gets replaced in a router's cache. A higher replacement rate indicates that the router frequently removes content from its cache, which increases the processing overhead. Additionally, a higher eviction rate at important routers can negatively impact the performance of caching techniques.\\
   \noindent \textit{9) Popularity Estimation:} It measures how often content is requested. If a content item reaches a certain threshold (predefined or dynamically estimated), it is considered popular. Popularity can be measured locally or globally, depending on caching needs. This measurement can be done either on-demand or periodically after a fixed time interval. \\
   \noindent \textit{10) Message Overhead :} Message overhead measures the extra messages needed to retrieve content. Overhead is incurred due to cooperation for caching and popularity estimation. This includes communication between controllers and routers, routers and community/cluster leaders, leaders with each other, routers with each other, and the retransmissions of Interests.

\subsection{Open Challenges and Future Research Directions} \label{challenges}

In-network caching is a key component of any ICN architecture. Making caching decisions remains one of the major challenges in ICN/NDN due to limited storage. Despite notable advancements in ICN/NDN caching technologies in recent years, there is still room for further investigation and improvement. This research area focuses on developing efficient caching techniques to minimize the overhead due to cooperation, optimizing the searching and placement of content within the neighborhood, and addressing challenges such as \emph{scalability, handling dynamic consumer demands, popularity estimation, limited cache capacity, content size, node mobility, energy constraints, the time involved in searching for content, storing content in the cache}, and \emph{removing content from the cache}, and more. In the following, we will look at these challenges and outline potential research directions to further improve in-network caching techniques performance in ICN/NDN.


\noindent \textit{1) Cooperative Caching:} Cooperative techniques are scalable and make better use of router storage located at both on-path and off-path (outside the transmission path). These techniques have been proven to improve the cache hit ratio and reduce content redundancy \cite{Reshadinezhad2023, Chaudhary2025}. However, a major challenge with cooperative techniques is establishing cooperation. Several techniques have been explored by researchers for establishing cooperation: 

(i) One solution is to use a centralized controller to establish communication among nodes \cite{Jmal2017a, Zhang2020a, Tavasoli2024}. However, the problem with this approach is the risk of a single point of failure and overloading of the central controller due to frequent changes in requests. 

(ii) Another solution is for each node to establish communication with its 1-hop, 2-hop, or N-hop neighbors \cite{Chaudhary2023, Hu2020a, Reshadinezhad2023}. Nevertheless, this is challenging due to the volatility of cached content, which results in frequent advertisement of requests across multiple interfaces, increasing communication overhead. 

(iii) A third approach involves dividing the network topology into smaller communities/clusters and selecting a leader/cluster head within each community to establish cooperation \cite{Huang2019, Gupta2021}. This method can help balance the load and reduce overhead. However, a problem arises when the selection of leaders is based solely on the connectivity of nodes, as this may not always be the best choice. In real-world scenarios, the storage and processing power of nodes are not uniform. Therefore, other characteristics of nodes, such as transmission cost, storage and processing capacity, and their location relative to other communities, should also be considered in the evaluation. Another aspect that needs exploration is how to divide the network topology into communities/clusters and determine the appropriate size of each cluster to balance the load among them.

\noindent \textit{2) Cache Replacement:} The role of the replacement policy is to make room for the new content when the cache is full. Most studies \cite{Amadeo2022, He2024, Li2024, Chaudhary2025} use LRU as the replacement policy to accommodate newly arrived content. The LRU policy has been mainly adopted due to its simplicity, as it does not cause overhead in replacement operations. However, its performance may not be optimal, so further studies are needed to explore alternative replacement techniques, especially under cooperative caching methods. In cooperative techniques, further exploration is required to determine which content to evict and whether caching evicted content in neighboring nodes is beneficial. Another area that needs attention is the real-world scenario where the size of each content may not be uniform. In such cases, how multiple chunks are evicted from the cache to accommodate new content needs further exploration.

\noindent \textit{3) Cache Diversity:} Caching a diverse range of contents within routers with limited capacity poses a significant challenge in ICN/NDN caching. When routers autonomously make decisions, it increases content redundancy by placing multiple copies of the same content in the network. Several prior works have introduced cooperative decision-making \cite{Chaudhary2023, Huang2019, Saha2013} to enhance cache diversity, but the question of how to establish this cooperation remains open. It can be achieved either through a centralized approach using an SDN controller with a complete view of the network topology or through routers coordinating with each other in a distributed manner.
Another challenge associated with increasing cache diversity is the potential for increased content retrieval time, especially when a single copy of the content is available in the network or domain \cite{Saino2013}. Exploring the tradeoff between cache diversity and content retrieval time is a key research direction in ICN.

\noindent \textit{4) Popularity Estimation:} Considering the limited cache capacity, popularity estimation is crucial for caching or replacement decisions. To assess the popularity of requested content, routers within the network maintain an additional table to monitor the flow of content requests. This popularity measurement can occur either locally \cite{Ong2014} or globally \cite{Gui2020}. In the former, each router autonomously manages a table to track passing requests, while in the latter, routers exchange local information to acquire the global request count from all consumers in the network. Given the substantial volume of requests from various consumer regions, global popularity proves beneficial for caching content at the network core. However, this approach introduces communication overhead as routers need to exchange information. 

Several key challenges associated with estimating content popularity at both local and global levels require further exploration. (i) Determining the optimal table size required to store information about previously requested content. (ii) How much weight should be given to local and global popularity? (iii) Which approach should be used to estimate popularity, whether on demand or at periodic intervals? The on-demand approach captures request patterns more accurately but may introduce overhead, while the periodic approach reduces overhead but is less effective when content request patterns change frequently. (iv) In periodic exchange, defining appropriate time intervals for updating the information in the table to reduce overhead and improve caching performance.

\noindent \textit{5) Content Popularity Model:} Most simulation-based studies \cite{Psaras2014, Wang2015, Amadeo2022, Li2024} model consumer requests using the Zipf distribution \cite{Breslau1999}, assuming that web requests follow a Zipf-like pattern. According to Zipf’s law, a few contents are highly popular and requested by the majority of consumers, while others are accessed less frequently. This assumption may be valid at a global level; however, at a regional level, content request patterns vary across different areas. Therefore, further investigation is needed to explore whether content popularity follows Zipf-like distribution at a local and global level by collecting real-world traces. Instead of restricting evaluation solely to the Zipf distribution, more research should involve collecting real-world web traces and performing time series analysis to better understand consumer demand patterns.

\noindent \textit{6) ICN/NDN Integration with SDN:} Integrating SDN and ICN improves caching performance in ICN by providing a full view of the network topology through the SDN controller. The SDN controller efficiently oversees available resources and installs rules to process ICN Interest and Data packets in the data plane. Due to the advantages of this solution, few prior works \cite{Nguyen2013a, Salsano2013, Zhang2020a, Tavasoli2024} have already proposed caching solutions by adopting SDN in the ICN/NDN environment. However, this integration presents several challenges, including scalability issues and the decision between implementing a single controller, susceptible to a single point of failure, or multiple controllers, prompting questions on how to establish cooperation. Moreover, deploying SDN in ICN presents a challenge, as OpenFlow switches, initially designed for processing IP packets, need to adapt to handling ICN packets. The integration of ICN with SDN is an active research area that needs sustained attention from researchers to utilize the benefits of these emerging technologies.

\noindent \textit{7) Evaluation Method:} Current research relies mainly on simulator-based evaluations \cite{Chaudhary2022, Khandaker2021, Wu2023}, usually focused on small-scale network topologies and assuming uniform capacity for each router and content size. However, it falls short of capturing real-world complexities. To address this gap, future work should involve examining the impact of caching techniques through a testbed evaluation method. The testbed evaluation method places network nodes across vast geographical areas with long distances between them. It uses actual hardware and software components that closely emulate the conditions and complexities observed in real-world networks. One can gain more reliable insights into performance, efficiency, and scalability by deploying caching mechanisms in the experimental testbed.

\noindent \textit{8) Determining the Practical Capacity of Routers:} In ICN/NDN, caching performance is closely tied to the storage capacity of routers. However, many existing caching techniques overlook the practical considerations of cache size \cite{Amadeo2022, Chaudhary2023, Psaras2012}. Therefore, a crucial area of investigation involves assessing the performance of caching mechanisms by determining the practical capacity of routers in real-world scenarios. This involves understanding how routers handle different-sized data requests and changing Internet traffic.

\noindent \textit{9) Hybrid Caching Solutions for ICN/NDN:} Both on-path and off-path methods possess their own set of advantages and disadvantages. The former minimizes overhead but suffers from performance issues \cite{Jacobson2009, Laoutaris2006, Psaras2012}, while the latter enhances performance but increases communication overhead \cite{Chaudhary2023, Hu2020a, Pal2021}. As a result, further studies should focus on designing hybrid caching solutions for ICN that achieve a balance between advertisement cost and gains in cache hit ratio. A hash-based technique \cite{Saino2013} is proven to effectively cache content in off-path routers, improving cache hit ratio and diversity. However, the distance between the cache node and the consumer can increase transmission costs. One way to solve this is by using a hybrid caching method combining hashing with cooperative techniques.

\section{Conclusion}

Information Centric Networking proposes a new communication model for the Internet to improve content delivery performance. 
In-network content caching is an inherent feature of these networks. 
Considering that many of these caching techniques have been designed recently, this remains an active area of research. In this article, we studied different caching techniques developed for ICN/NDN networks.
We first provide a taxonomy of ICN caching techniques by classifying them into eight categories: Centralized and Decentralized, On-Path and Off-Path, Popularity-Based, Probabilistic, Network-Centric, Content Granularity-Level, Freshness-Aware, and Machine/Deep Learning-Based. This paper then offers a comprehensive overview of the contributions made to caching in ICN/NDN so far.
We have also provided an overview of various relevant evaluation metrics used to assess the performance of these caching techniques, along with the challenges associated with deploying these caching techniques in real-world scenarios. Finally, we provided pointers to future work in this area highlighting the fact that both TCP/IP and ICN networks are likely to coexist in the near future.

\label{conclusion}

\bibliographystyle{ieee}
\bibliography{IEEEexample}

\end{document}